\documentclass[english,twocolumn]{revtex4-1}% Science two-column

\usepackage{graphicx}
\usepackage{bm}% bold math
\usepackage{amsmath}
\usepackage{amssymb}
\usepackage{verbatim}
\usepackage{color}
\usepackage{times}

\makeatletter
\usepackage[colorlinks=true, pdfstartview=FitV, linkcolor=blue, citecolor=blue, urlcolor=blue]{hyperref} % enable links

\def\H        {{$^1$H \/}}

\def\NN       {{$^{15}$N \/}}

\def\ie       {{\it i.e.}}

\newcommand{\ee}[1]{\cdot10^{#1}}
\newcommand{\mr}[1]{\mathrm{#1}}
\newcommand{\unit}[1]{\,\mathrm{#1}}

\newcommand{\us}{\,\mu{\rm s}}

\newcommand{\uHz}{\,\mu{\rm Hz}}
\newcommand{\ta}{t_\mr{a}}

\newcommand{\ts}{t_\mr{s}}
\newcommand{\tr}{t_\mr{r}}
\newcommand{\td}{t_\mr{d}}
\newcommand{\fac}{f_\mr{ac}}

\newcommand{\fc}{f_\mr{c}}
\newcommand{\fs}{f_\mr{s}}
\newcommand{\fAM}{f_\mr{AM}}

\newcommand{\fLI}{f_\mr{LI}}
\newcommand{\df}{\delta f}
\newcommand{\Df}{\gamma_\mr{int}}
\newcommand{\Cr}{C}
\newcommand{\Cthresh}{C_\mr{thresh}}

\newcommand{\ytrace}{\{y_k\}}
\newcommand{\sigY}{\sigma_Y}
\newcommand{\phimax}{\phi_\mr{max}}
\newcommand{\SNR}{\mr{SNR}}

\@booleantrue\preprintsty@sw
\def\frontmatter@abstractheading{\vspace{0.5cm}}
\g@addto@macro\abstract{\ignorespaces}

\renewcommand\@biblabel[1]{(#1)}

\renewcommand{\cite}[1]{{\it\citep{#1}}}
\setcitestyle{round}

\makeatother

\begin{document}

% formatting
\global\emergencystretch = .1\hsize % adjust the line-breaking to avoid overfull \hbox (standard .15\hsize)

\title{Quantum sensing with arbitrary frequency resolution}

\author{J. M. Boss$^\dagger$, K. S. Cujia$^\dagger$, J. Zopes, and C. L. Degen$^1$}
\email{degenc@ethz.ch} 
\thanks{$^\dagger$These authors contributed equally to this work.}
\affiliation{
$^1$Department of Physics, ETH Zurich, Otto Stern Weg 1, 8093 Zurich, Switzerland.
}
\date{\today}

\begin{abstract}
Quantum sensing takes advantage of well controlled quantum systems for performing measurements with high sensitivity and precision.  We have implemented a concept for quantum sensing with arbitrary frequency resolution, independent of the qubit probe and limited only by the stability of an external synchronization clock.  Our concept makes use of quantum lock-in detection to continuously probe a signal of interest.  Using the electronic spin of a single nitrogen vacancy center in diamond, we demonstrate detection of oscillating magnetic fields with a frequency resolution of $70\unit{\uHz}$ over a MHz bandwidth.  The continuous sampling further guarantees an excellent sensitivity, reaching a signal-to-noise ratio in excess of $10^4$ for a $170\unit{nT}$ test signal measured during a one-hour interval.  Our technique has applications in magnetic resonance spectroscopy, quantum simulation, and sensitive signal detection.
\end{abstract}

\maketitle

%\tableofcontents

%%%%% One-sentence summary

%\noindent{\large\bf One sentence summary}\vspace{0.5cm}

%\noindent Stroboscopic sampling allows a qubit sensor to resolve minute frequency variations with high sensitivity.

%\vspace{1cm}

%%%%% introduction

%\noindent{\large\bf Main text}\vspace{0.5cm}

%\noindent
Quantum sensors with new capabilities are driving the field of precision metrology \cite{giovannetti11,degen16}.  In particular, spin qubits associated with crystal defects in diamond \cite{doherty13} and other materials \cite{pla12,widmann15,kolesov12} have emerged as excellent probes with nanometer spatial resolution \cite{balasubramanian08,maletinsky12}.  Because the defect spins are well isolated from the environment, they can be controlled with high fidelity, allowing researchers to implement sophisticated quantum manipulation protocols.

A particularly important sensing task is the spectral decomposition of time-varying signals into their frequency components.  Quantum metrology employs techniques from quantum control to reach this goal.  For example, dynamical decoupling methods -- originally developed for protecting qubits from decoherence -- have been adapted for detecting alternating signals with narrow bandwidth and high signal-to-noise ratio \cite{cywinski08,kotler11,delange11,alvarez11}.  Other, more recent techniques include dressed-state approaches \cite{yan13,loretz13}, Floquet spectroscopy \cite{lang15}, and correlative measurements \cite{laraoui10,laraoui13}.  Crucially, the spectral resolution of all of these techniques is limited by the state life time of the qubit probe.  For nitrogen vacancy centers in diamond, reported spectral resolutions are a few Hz at best, even when assisted by a long-lived quantum memory \cite{zaiser16,rosskopf16,pfender16}.

We introduce a simple concept where the frequency estimation is solely limited by the stability of an external, classical reference clock and the total available measurement time.  Our method takes advantage of the quantum lock-in amplifier \cite{kotler11,delange11} which is used to stroboscopically sample the signal of interest.  Although the acquired signal is highly undersampled, we show that the original wideband spectrum can be recovered by compressive sampling methods \cite{donoho06}. The periodic sampling further guarantees that the signal-to-noise ratio (SNR) increases in proportion to the measurement time.  We demonstrate our method by recording signal traces for up to $4\unit{hours}$, reaching a frequency resolution of $70\unit{\uHz}$ and a precision of $260\unit{nHz}$ with $\SNR>10^4$.
%$\sim 3.5\times10^4$.

Our experimental demonstration makes use of a spin qubit formed by the negatively-charged nitrogen vacancy (NV) center in diamond. The NV center is a suitable object for our demonstration because it can be efficiently initialized, manipulated and read out at room temperature by optical and microwave pulses.  Furthermore, its sensing technology is well developed \cite{maletinsky12} and addresses a broad range of potential applications in physics, materials science and biology \cite{rondin14,schirhagl14}.
\begin{figure*}
%\centering
\includegraphics[width=0.90\textwidth]{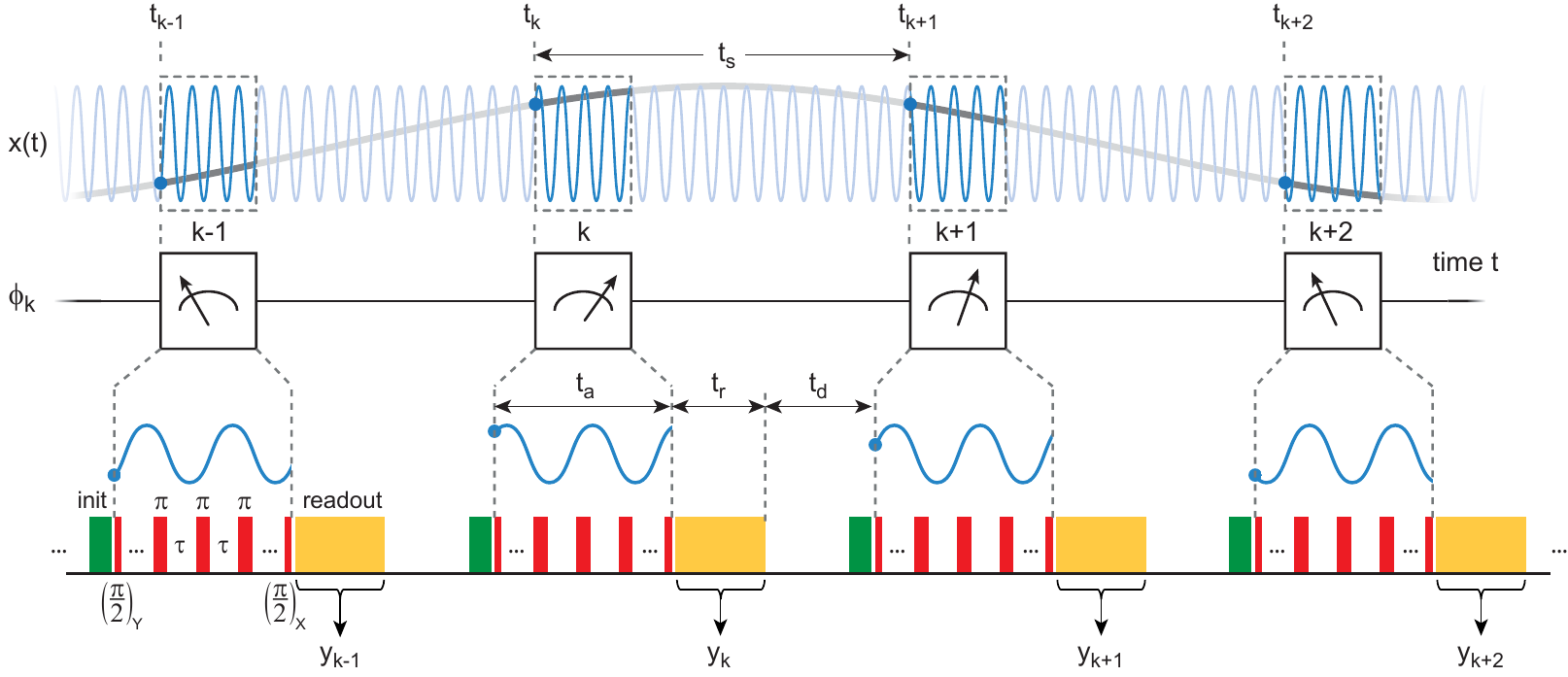}
\caption{
Basic concept of continuous sampling.  The qubit sensor stroboscopically probes an a.c. signal $x(t)$ in intervals of the sampling period $\ts$.  Each sampling instance $k$ consists of sensor initialization (green), a phase measurement using quantum lock-in detection (red pulses), and sensor readout (yellow).  A sensor output $y_k$ is proportional to the quantum phase $\phi_k$ and to the instantaneous value of $x(t_k)$ at time stamp $t_k$ (blue dots).  A time trace $\ytrace$ of sensor outputs therefore contains the undersampled signal $x(t_k)$ (grey oscillation).
In our experiment, the sensor qubit was implemented by the electronic spin of a nitrogen-vacancy center in diamond.  A laser pulse was used for initialization and nuclear-spin-assisted optical detection for readout.
Quantum lock-in detection was implemented by a CPMG sequence with interpulse spacing $\tau$ and either 16 or 32 $\pi$ pulses (see Materials and Methods).
%XY8 phase cycling scheme \cite{gullion90} 
%corresponding to $\ta = 6-26\unit{\us}$.
}
\label{fig:fig1}
\end{figure*}
%

%%% Theory/Concept

Our approach (Fig. \ref{fig:fig1}) relies on periodic sampling of a signal $x(t)$ in intervals of a sampling period $\ts$.  Each sampling instance consists of three periods, including a quantum lock-in measurement of duration $\ta$, qubit state readout during $\tr$, and an additional delay time $\td$ to accommodate for experimental overhead and to adjust the sampling rate.  The sampling period is then $\ts=\ta+\tr+\td$.

To implement the quantum lock-in measurement, we use a Carr-Purcell-Meiboom-Gill (CPMG) decoupling sequence (red pulses in Fig. \ref{fig:fig1}).  Specifically, we initialize the qubit to the $+X$ state of the $X$ basis and modulate it by a series of $\pi$ pulses with inter-pulse spacing $\tau$.  This defines the lock-in detection frequency $\fLI=m/(2\tau)$ where $m=1,3,5,...$ is the harmonic order.  The frequency bandwidth of the lock-in is approximately $\fLI\pm 1/(2\ta)$ \cite{delange11,degen16}.  For an a.c. signal $x(t) = \Omega\cos(2\pi\fac t)$ with a frequency $\fac\approx\fLI$ within this bandwidth, the quantum phase accumulated after time $\ta$ is
\begin{equation}
\phi_k = \frac{2\ta}{\pi}x(t_k) \ , \label{eq:phase}
\end{equation}
where $t_k$ marks the start of the lock-in measurement and $\Omega$ is the signal amplitude in units of angular frequency (Supplementary Text 1).  Crucially, although the quantum phase is accumulated over an extended time interval $[t_k,t_k+\ta]$, its value reflects the instantaneous value of $x(t)$ at time $t=t_k$. 
To read out the quantum phase, the quantum state is measured in the $Y$ basis yielding a probability
\begin{equation}
p_k = \frac12(1-\sin\phi_k) \approx \frac12(1-\phi_k) \label{eq:probability}
\end{equation}
to find the system pointing along the $-Y$ direction.  The approximation is for small $|\phi_k|\ll \pi/2$ within the sensor's linear range \cite{kotler13}.  Optical readout finally converts the projected state into a photon number $y_k$.  Because state projection and optical readout are stochastic processes, $y_k$ is a random variable,
\begin{equation}
y_k = \mathrm{Pois} \left[ \Cr (1 - \epsilon\,\mathrm{Bn}[p_k] )\right] \ , \label{eq:readout}
\end{equation}
where \textit{Bn} is a Bernoulli process which takes the value 1 with a probability of $p_k$ and the value 0 with probability $1-p_k$, and \textit{Pois} is a Poisson process that reflects the photon shot noise.  $\Cr$ is a variable readout gain and $\epsilon$ is the optical contrast.

By collecting a time trace of $N$ measurement outputs $\ytrace_{k=1}^{N}$ at sampling times $t_k = k\ts$, we can sample the signal $x(t)$ at a sub-Nyquist rate $\fs=1/\ts$.  Hence, a Fourier transform of the time trace reveals a discrete undersampled spectrum of $x(t)$.  Crucially, the number of samples $N$ can be made as large as desired, allowing for a frequency resolution $\df=\fs/N$ that is arbitrarily fine.

%%%%% experimental

To implement our continuous sampling technique, we used the qubit formed by the $m_S=0$ and $m_S=-1$ spin sub-levels of single NV centers located in diamond nanopillar waveguides (see Materials and Methods).  At a bias field of $457\unit{mT}$ applied along the NV symmetry axis, the transition frequency between these states is $9916\unit{MHz}$.  We initialized the qubit using a 532~nm laser pulse and a microwave $\pi/2$ pulse, and detected the qubit state using a phase-shifted $\pi/2$ pulse followed by optical readout.
%We initialized the qubit into the $X$ basis using a 532~nm laser pulse and a microwave $\pi/2$ pulse, and detected the qubit state in the $Y$ basis using a phase-shifted $\pi/2$ pulse followed by optical readout.
%
We used an indirect readout scheme where the final qubit state was first stored in the \NN nuclear spin ($I=1/2$), serving as a memory qubit \cite{rosskopf16}, and we then repetitively read out the \NN spin state by a nuclear quantum non-demolition (QND) measurement \cite{jiang09,neumann10science}.  By varying the number of QND measurements $n$ we could adjust the readout gain $\Cr$ between ca. $0-230$ photons.  The optical contrast was $\epsilon\approx 0.35$.
%  The optical readout gain was $\Cr \approx n\times 0.15\unit{photons}$ with $\tr = n\times 2.32\unit{\us}$, and the optical contrast was $\epsilon\approx 0.33$.
%$\Cr\approx n\times 0.11\unit{photons}$
%$\tr = n\times 2.32\unit{\us}$.
% \approx \tr 47\unit{photons/ms}$.

%%%%% RESULTS

\begin{figure}
%\centering
\includegraphics[width=0.45\textwidth]{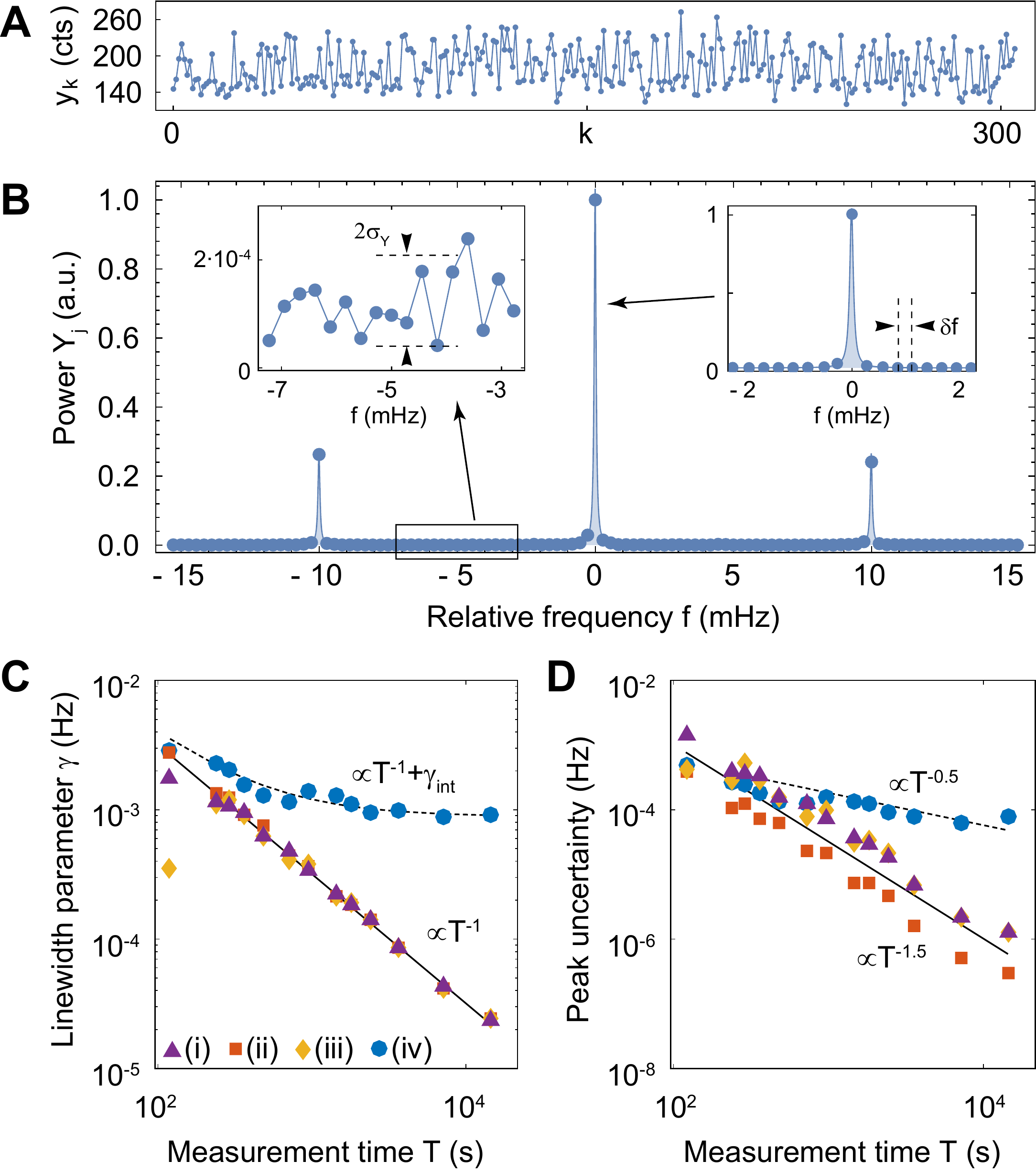}
\caption{
Continuous sampling of a.c. magnetic signals.
({\bf A}) $1.26\unit{s}$ excerpt of a one-hour time trace sampled at $\ts=4.2\unit{ms}$.  Vertical axis shows photon counts.  The telegraph-like behavior results from the stochastic quantum state projection.
({\bf B}) Fourier spectrum (power) of the full one-hour time trace.  Three peaks are visible that correspond to the central and side peaks of the amplitude-modulated signal.  The horizontal axis indicates the detuning from the carrier frequency.  The insets show the noise floor (left inset) and the frequency resolution (right inset) of the spectrum.
({\bf C}) Fitted linewidth paramater $\gamma$ as a function of total measurement time $T$ for four signals. Datapoints (i-iii) originate from coherent signals and datapoints (iv) from an incoherent signal with an artificial line broadening of $\Df = 0.76\unit{mHz}$.
Original spectra are given in Fig. S3.
%({\bf C})  Power spectrum of a.c. signals for $T=4\unit{min}$, $10\unit{min}$ and $240\unit{min}$.  Peaks (i-iii) originate from coherent signals and peak (iv) originates from an incoherent signal with an artificial line broadening of $\Df = 0.76\unit{mHz}$.
%({\bf D}) Fitted linewidth paramater $\gamma$ for the different peaks (refer to C for color coding).
({\bf D}) Uncertainty in the fitted peak frequencies for the different peaks.  Solid and dashed lines are guides to the eye.
}
\label{fig:fig2}
\end{figure}
As a first illustration of the continuous sampling technique, Figs.~\ref{fig:fig2}A,B show a time trace and spectrum of an amplitude modulated (AM) magnetic test signal with carrier frequency $\fc=601.2547\unit{kHz}$ and modulation frequency $\fAM = 10\unit{mHz}$.  The test signal had an amplitude of approximately $170\unit{nT}$, corresponding to $\Omega = 2\pi\times 4.7\unit{kHz}$, and was generated by passing an a.c. current through a nearby wire.  The signal contained three components at frequencies $\fc$ and $\fc \pm \fAM$ with a power ratio of 1:4:1.  The frequency resolution $\df$ of the spectrum, obtained from a time trace of one hour duration, was $\df=1/T = 278\unit{\uHz}$ (Fig.~\ref{fig:fig2}B, right inset).  Because our signal was undersampled, the abscissa in Fig.~\ref{fig:fig2}B indicates the detuning from $\fc$ rather than the absolute frequency.  The observed $\uHz$ frequency resolution and the consistent amplitude ratio between carrier and side peaks illustrate the capabilities of our method.

While our strategy allows for an arbitrary frequency resolution $\delta f$, we are more interested in how precisely we can determine a signal's linewidth and center frequency in the experiment.  Fig.~\ref{fig:fig2}D depicts the decrease in the fitted linewidth parameter $\gamma$ with increasing measurement time $T$ for four signals.  Signals (i-iii) were produced by amplitude modulation similar to Fig.~\ref{fig:fig2}B and had zero intrinsic linewidth, $\Df=0$.  The linewidth parameter for these signals scaled with $\gamma\propto T^{-1}$, which represents the Fourier transform limit of the continuous sampling method.  The $T^{-1}$ scaling is expected to continue until the phase noise in the reference clock or the frequency jitter in the signal generator become dominating.  Signal (iv), on the other hand, was artificially broadened by frequency modulation (FM) with Gaussian noise in order to mimic a non-zero intrinsic linewidth $\Df>0$.  The linewidth parameter for signal (iv) initially also decreased as $T^{-1}$ but leveled out as $\gamma$ approached $\Df$.  Fig.~\ref{fig:fig2}D further shows the fit errors in the peak frequencies for all signals.  Here, a $T^{-1.5}$ scaling was observed that reduced to $T^{-0.5}$ once the intrinsic linewidth became significant, as expected for the scaling of the spectral amplitude variance (Supplementary Text 2).

%While our strategy allows for an arbitrary frequency resolution $\delta f$, we are more interested in how precisely we can determine a signal's linewidth and center frequency in the experiment.  Fig.~\ref{fig:fig2}C depicts the decrease in the linewidth parameter $\gamma$ with increasing measurement time $T$ for four signals.  Signals (i-iii) were coherent signals at $\fc=1.2\unit{MHz}$ and $\fc\pm 15\unit{mHz}$ produced by amplitude modulation.  Signal (iv) at $\fc + 40\unit{mHz}$ was artificially broadened by frequency modulation (FM) with Gaussian noise in order to mimic an intrinsic signal linewidth.  As expected, the linewidth parameters $\gamma$ of the coherent signals (i-iii) scaled with $T^{-1}$ (Fig.~\ref{fig:fig2}D) and reached values around $44\unit{\uHz}$ for the maximum duration of $T=4\unit{h}$.  This scaling is expected to continue until the phase noise in our reference clock and the frequency jitter between signal generators is reached.  By contrast, the linewidth parameter of the broadened signal (iv) is limited by the FM modulation amplitude $\Df=0.76\unit{mHz}$.  Fig.~\ref{fig:fig2}E further shows the fit error in the peak frequencies for all signals.  In this case both signals initially followed a $T^{-1.5}$ scaling with total measurement duration $T$.  The scaling flattened to $T^{-0.5}$ for the FM signal once the intrinsic linewidth was reached, as expected for the scaling of the spectral amplitude variance (Supplementary Text 2).

%%%%% SNR

\begin{figure}[b!]
%\centering
\includegraphics[width=0.45\textwidth]{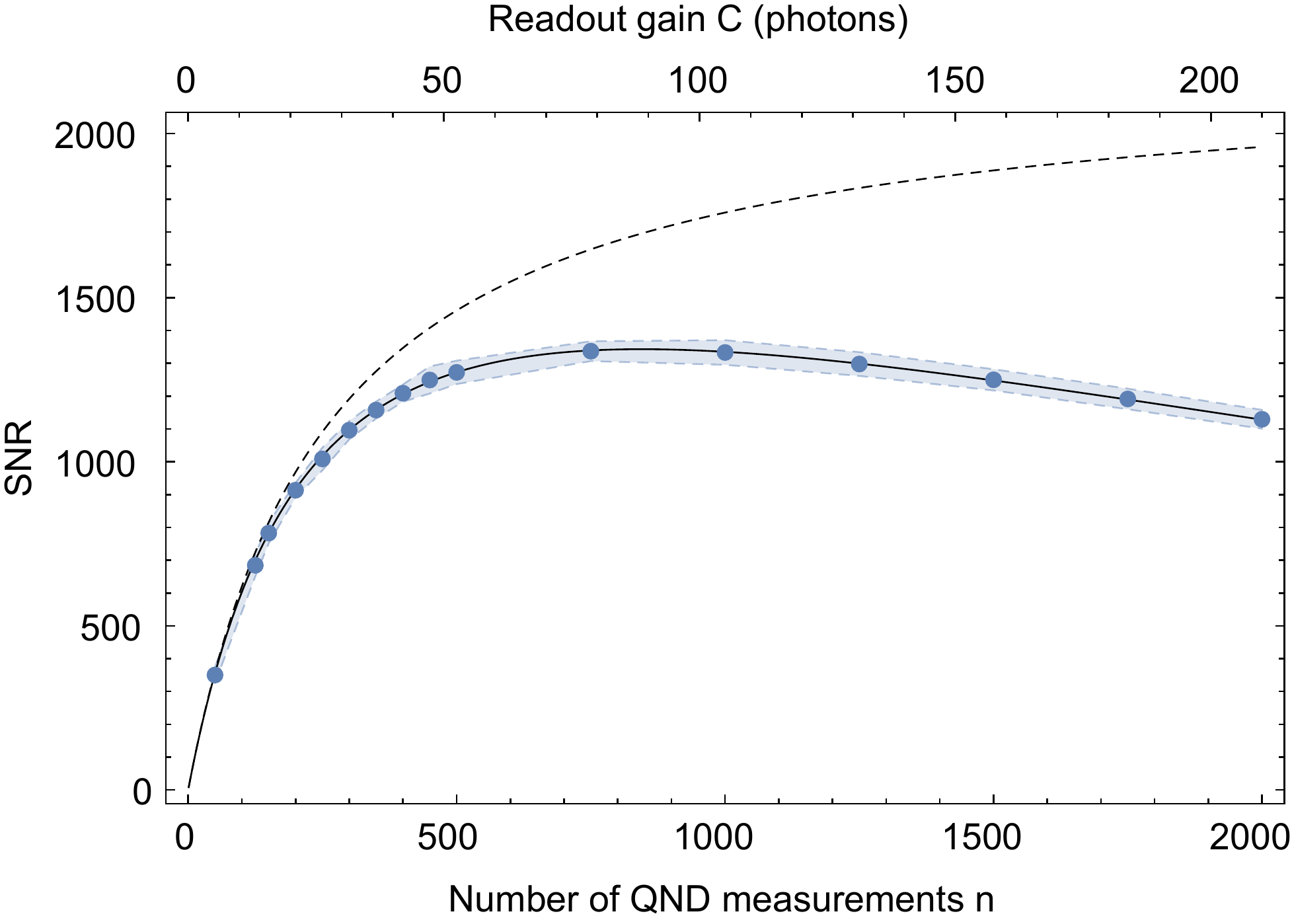}
\caption{
Signal-to-noise ratio (SNR) as a function of the readout gain $C$.  The readout gain is adjusted via the number of repetitive QND measurements $n$ of the nuclear memory qubit.  Blue dots are the experimental data and represent the mean SNR of six time traces with $N=3.85\ee{5}$ samples.  The dashed line represents the ideal SNR predicted by theory (Eq. \ref{eq:snr}).  The solid line in addition takes  the depolarization of the nuclear \NN memory qubit into account (Supplementary Text 3).  The filled blue region indicates the standard error of the fit.
From our data we extract a \NN depolarization rate $\Gamma\approx 1.4\ee{-4}\mr{/readout}$.
%From our data we extract a readout gain $\Cr(n)=n\times 0.105\unit{photons}$, an optical contrast $\epsilon=0.35$ and a \NN depolarization rate  $\Gamma\approx 1.4\ee{-4}\mr{/readout}$.
The corresponding optimum SNR ($n=260$) for a one-hour measurement interval is $1.2\ee{4}$ (Materials and Methods).
}
\label{fig:fig3}
\end{figure}
We next examined how the sensitivity of the sensor can be optimized.  In order to quantify the sensitivity, we compared the peak amplitude $Y_j$ with the standard deviation $\sigY$ of the noise floor in the power spectrum (see Fig \ref{fig:fig2}B).
This defines a power signal-to-noise ratio,
\begin{align}
\mr{SNR} = \frac{Y_j} {\sigY} \ .
\end{align}
Assuming that the entire signal power is concentrated in a single Fourier component $Y_j$ of the spectrum (\ie, that the linewidth of the signal is smaller than $\df$), it follows from Eq. \ref{eq:probability} and \ref{eq:readout} that
\begin{align}
Y_j \approx \frac{1}{16} N^2\Cr^2\epsilon^2\phimax^2  \ ,
\label{eq:signal}
\end{align}
where $\phimax=2\ta\Omega/\pi$ is the signal amplitude expressed in units of the accumulated phase (Supplementary Text 3).  The approximation is again for small signals within the linear response of the lock-in (Eq. \ref{eq:probability}).  The noise $\sigY$ is the sum of two contributions, one from quantum projection noise with variance $\frac14 N\Cr^2\epsilon^2$ and one from optical shot noise with variance $N\Cr(1-\frac{\epsilon}{2})$.  The SNR becomes
\begin{align}
\mr{SNR} = \frac{\frac{1}{16}N\Cr^2\epsilon^2\phimax^2}{\frac14\Cr^2\epsilon^2+\Cr(1-\frac{\epsilon}{2})}   \ . \label{eq:snr}
\end{align}
Because $N=T\fs$, the SNR improves proportional to the duration of the time record $T$.

To further optimize the SNR, we adjusted the phase amplitude $\phimax$ and the readout gain $\Cr$.  We achieved this by varying the sensing time $\ta$ and the readout time $\tr$.  First, we increased the sensing time $\ta$ so that the quantum phase covered the full linear range of the lock-in, typically $\phimax \sim 0.5$.  Although larger $\phimax$ are possible, the response of the lock-in becomes non-linear \cite{kotler13} and harmonics are generated in the spectrum.  This complicates the interpretation while providing little further improvement in the SNR (Supplementary Text 1 and Figs. S4, S5).
Next, we turned up the gain $\Cr$ until sensor readout became dominated by quantum projection noise.  In our experiment, we could adjust $\Cr$ by varying the number $n$ of QND measurements of the nuclear memory qubit, where $C(n) \approx n \times 0.105\unit{photons}$ and $\tr \approx n \times 2.32\unit{\us}$.
Fig.~\ref{fig:fig3} plots the SNR for signal (ii) in Fig.~\ref{fig:fig2}C as a function of $n$.  The SNR increased rapidly for small $n$ until it saturated around $n\approx 260$, which corresponded to the threshold gain $\Cthresh = 4/\epsilon^2-2/\epsilon \approx 27$ where shot noise and quantum projection noise are balanced \cite{degen16}.  Increasing the gain beyond $\Cthresh$ only marginally improved the SNR and eventually even degraded it.  The degradation at very high gains was due to the imperfection of the nuclear quantum memory, which became depolarized under optical illumination \cite{neumann10science}.
% where $\Cr\approx n\times 0.11\unit{photons}$ and $\tr = n\times 2.32\unit{\us}$.
% \approx \tr 47\unit{photons/ms}$.

%%%%% compressive sampling

\begin{figure}
%\centering
\includegraphics[width=0.45\textwidth]{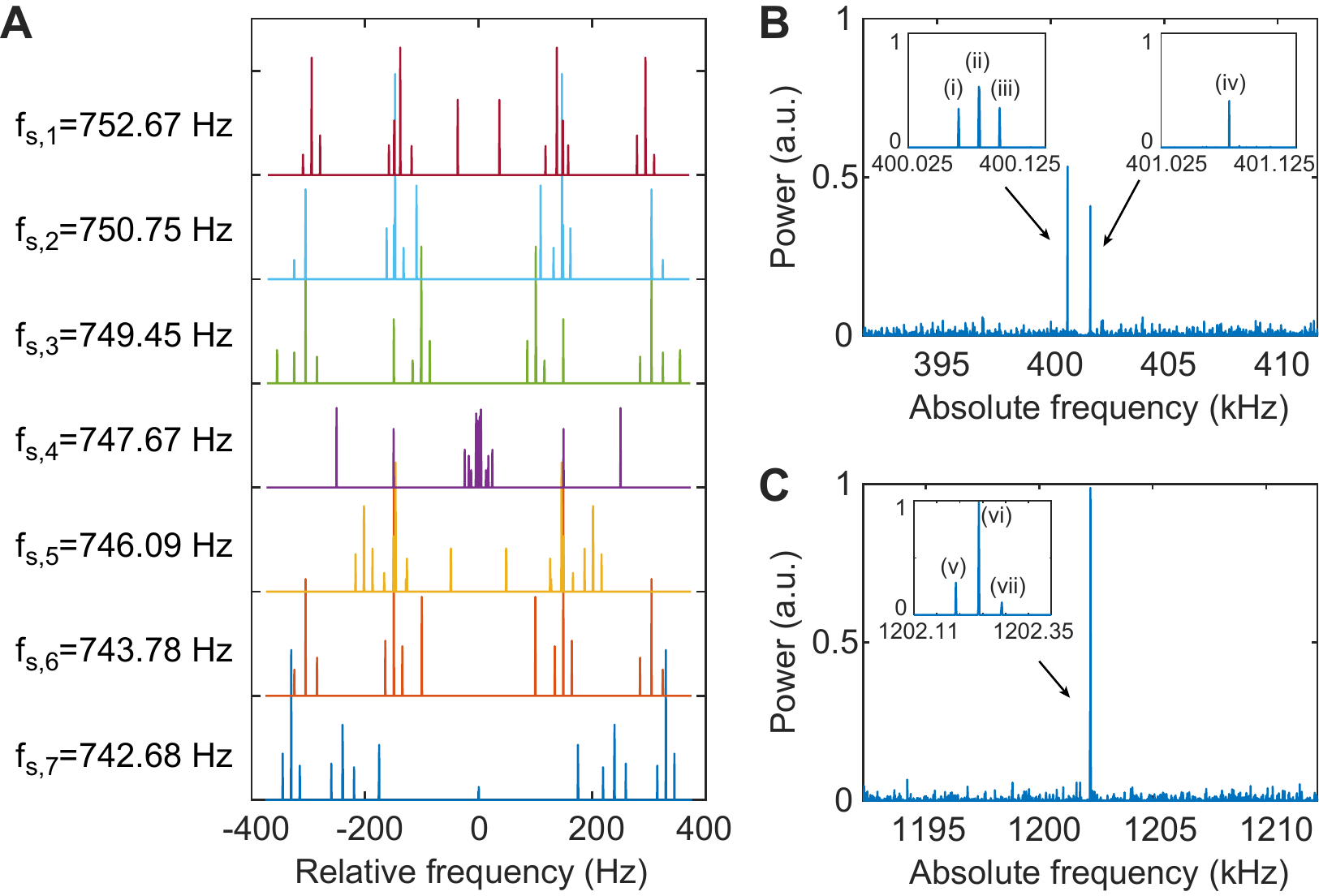}
\caption{
Wideband spectral reconstruction based on compressive sampling.
({\bf A}) Undersampled spectra obtained from time records with $i=1...7$ random chosen sampling rates $f_\mr{s,i}$.  The data represent records of duration $T=2\unit{s}$ that were averaged for $1.5$ hours.
({\bf B,C}) Reconstructed wideband spectrum containing $l=7$ tones in two sub-bands.  The insets show the individual tones with frequencies $f^\mr{(i)}=\fc-15\unit{Hz}$, $f^\mr{(ii)}=\fc$, $f^\mr{(iii)}=\fc+15\unit{Hz}$, $f^\mr{(iv)}=\fc+1000\unit{Hz}$ (where $\fc=400.75\unit{kHz}$), and $f^\mr{(v)}=\fc'-40\unit{Hz}$, $f^\mr{(vi)}=\fc'$, $f^\mr{(vii)}=\fc'+40\unit{Hz}$ (where $\fc'=1202.25\unit{kHz}$).  The noise reflects the incomplete image rejection of the reconstruction procedure.
}
\label{fig:fig4}
\end{figure}
Thus far, all our measurements reported relative rather than absolute signal frequencies.  The measurement of absolute signal frequencies is hindered by the large undersampling.  For example, in Fig.~\ref{fig:fig2}B, a signal of frequency $\fc \sim 601\unit{kHz}$ was sampled at $\fs = 0.237\unit{kHz}$, which is about $5\ee{3}$ times slower than the Nyquist rate.  We now discuss a strategy that overcomes this limitation using compressive sampling. 
We implemented this strategy by recording a set of time traces with slightly different sampling rates $\fs$.
%f1=601.2547 kHz and fs = 237.444 Hz, 5064 times lower than the Nyquist rate

Compressive sampling (CS) exploits our prior knowledge about the sparsity of the wideband spectrum \cite{donoho06,candes06}.  Suppose the vector $\vec{X}$ holds the Fourier components of the desired wideband spectrum sampled at or above the Nyquist rate, and the vectors $\vec{Y}_i$ represent a small set of undersampled spectra with $i=1,2,\dots,p$.  We can express our measured undersampled spectra $\vec{Y}_i$ by the linear system 
\begin{align}
\begin{pmatrix}\vec{Y}_1 \\ \vec{Y}_2\\ \vdots \\ \vec{Y}_p\end{pmatrix}
= \begin{pmatrix}\Phi_1 \\ \Phi_2\\ \vdots \\ \Phi_p\end{pmatrix}  \vec{X} \ ,
%=  \mathbf{\Phi}  \vec{X}.
 \label{eq:compressiveproblem1}
\end{align}
where $\Phi_i$ are sampling matrices folding the wideband spectrum into the bandwidths of the undersampled spectra \cite{sun12}.  To reconstruct the wideband spectrum, we solve Eq. (\ref{eq:compressiveproblem1}) for $\vec{X}$.  Although the linear system is highly underdetermined, a solution can be found if $\vec{X}$ is sparse (is significantly non-zero only for a few frequencies) and the $\Phi_i$ are mutually incoherent.

To demonstrate wideband spectral reconstruction, we implemented a CS scheme to recover $l=7$ tones from a set of $p=7$ undersampled spectra (Supplementary Text 4).  We adjusted the sampling frequencies $\fs$ via the delay time $\td$.  To ensure incoherence between the sampling matrices, we randomized our choices of $\td$.
%We solved Eq. (\ref{eq:compressiveproblem1}) \textcolor{red}{using a non-negative least squares solver [cite]}.
Fig.~\ref{fig:fig4} shows the undersampled spectra together with the reconstructed wideband spectrum.  The tones in these spectra were contained in two 20~kHz-wide frequency bands, one centered at the first harmonic of the lock-in filter function at 400~kHz and one at the third harmonic at 1,200~kHz.  Although the SNR of the reconstructed spectrum is reduced due to incomplete image rejection, the experiment clearly demonstrates that the absolute peak frequencies can be unambiguously recovered.  The image rejection can be improved by increasing the number of spectra $p$.

%%%%% Outlook

Our experiments demonstrate that a quantum sensor can achieve a frequency resolution far beyond its intrinsic state lifetime, limited only by the stability of an external synchronization clock.  Looking forward, quantum sensing with arbitrary frequency resolution has important applications in sensitive magnetic and electric field detection.  A high spectral resolution is for example essential for nanoscale nuclear magnetic resonance (NMR) imaging experiments \cite{mamin13,staudacher13,loretz14apl}, where minute spectral shifts can be used to infer atomic positions, internuclear distance vectors, and molecular connectivity.
Spectral addressability is also important for operating large-scale quantum registers in solid-state quantum simulators \cite{cai13}.
Although NMR spectra are often broadened by internuclear interactions, a rich repertoire of line narrowing and isotope dilution techniques exists for refining the spectral resolution \cite{mehring12,reif12}.  Ultrahigh resolution NMR is able to resolve couplings of a few mHz \cite{appelt06} under favorable conditions and achieves $<20\unit{Hz}$ linewidths even for \H in dense solid samples \cite{reif12}.
Finally, continuous sampling can provide sensitivity gains when measuring weak, modulated signals.  This is because of the high duty cycle achieved by continuously probing the signal during the measurement time $T$, combined with the favorable $\propto T$ scaling of the SNR.
%Applications include quantum imaging of single spins \cite{grinolds13} or current flow \cite{chang17} in nanoscale structures.

%%%%% Acknowledgments

%\newpage
%\noindent{\large\bf Acknowledgements}\vspace{0.5cm}

%\noindent
We thank T. Rosskopf, K. Chang, A. Retzker and F. Jelezko for experimental support and useful discussions. This work was supported by the Swiss NSF Project Grant $200021\_137520$, by the Swiss NSF NCCR QSIT, and by the FP7-611143 DIADEMS programme of the European Commission.

%%%%% arXiv

\textit{Note added to the arXiv version:}
This arXiv preprint corresponds to the manuscript resubmitted to \textit{Science} \cite{boss17} with some further literature references included.
We also acknowledge similar work by Jelezko and coworkers \cite{schmitt17} and Walsworth and coworkers \cite{bucher17}.


\begin{thebibliography}{38}%
\makeatletter
\providecommand \@ifxundefined [1]{%
 \@ifx{#1\undefined}
}%
\providecommand \@ifnum [1]{%
 \ifnum #1\expandafter \@firstoftwo
 \else \expandafter \@secondoftwo
 \fi
}%
\providecommand \@ifx [1]{%
 \ifx #1\expandafter \@firstoftwo
 \else \expandafter \@secondoftwo
 \fi
}%
\providecommand \natexlab [1]{#1}%
\providecommand \enquote  [1]{``#1''}%
\providecommand \bibnamefont  [1]{#1}%
\providecommand \bibfnamefont [1]{#1}%
\providecommand \citenamefont [1]{#1}%
\providecommand \href@noop [0]{\@secondoftwo}%
\providecommand \href [0]{\begingroup \@sanitize@url \@href}%
\providecommand \@href[1]{\@@startlink{#1}\@@href}%
\providecommand \@@href[1]{\endgroup#1\@@endlink}%
\providecommand \@sanitize@url [0]{\catcode `\\12\catcode `\$12\catcode
  `\&12\catcode `\#12\catcode `\^12\catcode `\_12\catcode `\%12\relax}%
\providecommand \@@startlink[1]{}%
\providecommand \@@endlink[0]{}%
\providecommand \url  [0]{\begingroup\@sanitize@url \@url }%
\providecommand \@url [1]{\endgroup\@href {#1}{\urlprefix }}%
\providecommand \urlprefix  [0]{URL }%
\providecommand \Eprint [0]{\href }%
\providecommand \doibase [0]{http://dx.doi.org/}%
\providecommand \selectlanguage [0]{\@gobble}%
\providecommand \bibinfo  [0]{\@secondoftwo}%
\providecommand \bibfield  [0]{\@secondoftwo}%
\providecommand \translation [1]{[#1]}%
\providecommand \BibitemOpen [0]{}%
\providecommand \bibitemStop [0]{}%
\providecommand \bibitemNoStop [0]{.\EOS\space}%
\providecommand \EOS [0]{\spacefactor3000\relax}%
\providecommand \BibitemShut  [1]{\csname bibitem#1\endcsname}%
\let\auto@bib@innerbib\@empty
%</preamble>
\bibitem [{\citenamefont {Giovannetti}\ \emph {et~al.}(2011)\citenamefont
  {Giovannetti}, \citenamefont {Lloyd},\ and\ \citenamefont
  {Maccone}}]{giovannetti11}%
  \BibitemOpen
  \bibfield  {author} {\bibinfo {author} {\bibfnamefont {V.}~\bibnamefont
  {Giovannetti}}, \bibinfo {author} {\bibfnamefont {S.}~\bibnamefont {Lloyd}},
  \ and\ \bibinfo {author} {\bibfnamefont {L.}~\bibnamefont {Maccone}},\ }\href
  {\doibase 10.1038/NPHOTON.2011.35} {\bibfield  {journal} {\bibinfo  {journal}
  {Nature Photonics}\ }\textbf {\bibinfo {volume} {5}},\ \bibinfo {pages} {222}
  (\bibinfo {year} {2011})}\BibitemShut {NoStop}%
\bibitem [{\citenamefont {Degen}\ \emph {et~al.}(2016)\citenamefont {Degen},
  \citenamefont {Reinhard},\ and\ \citenamefont {Cappellaro}}]{degen16}%
  \BibitemOpen
  \bibfield  {author} {\bibinfo {author} {\bibfnamefont {C.~L.}\ \bibnamefont
  {Degen}}, \bibinfo {author} {\bibfnamefont {F.}~\bibnamefont {Reinhard}}, \
  and\ \bibinfo {author} {\bibfnamefont {P.}~\bibnamefont {Cappellaro}},\
  }\href {https://arxiv.org/abs/1611.02427} {\bibfield  {journal} {\bibinfo
  {journal} {arXiv:1611.02427}\ } (\bibinfo {year} {2016})}\BibitemShut
  {NoStop}%
\bibitem [{\citenamefont {Doherty}\ \emph {et~al.}(2013)\citenamefont
  {Doherty}, \citenamefont {Manson}, \citenamefont {Delaney}, \citenamefont
  {Jelezko}, \citenamefont {Wrachtrup},\ and\ \citenamefont
  {Hollenberg}}]{doherty13}%
  \BibitemOpen
  \bibfield  {author} {\bibinfo {author} {\bibfnamefont {M.~W.}\ \bibnamefont
  {Doherty}}, \bibinfo {author} {\bibfnamefont {N.~B.}\ \bibnamefont {Manson}},
  \bibinfo {author} {\bibfnamefont {P.}~\bibnamefont {Delaney}}, \bibinfo
  {author} {\bibfnamefont {F.}~\bibnamefont {Jelezko}}, \bibinfo {author}
  {\bibfnamefont {J.}~\bibnamefont {Wrachtrup}}, \ and\ \bibinfo {author}
  {\bibfnamefont {L.~C.}\ \bibnamefont {Hollenberg}},\ }\href
  {http://www.sciencedirect.com/science/article/pii/S0370157313000562}
  {\bibfield  {journal} {\bibinfo  {journal} {Physics Reports}\ }\textbf
  {\bibinfo {volume} {528}},\ \bibinfo {pages} {1} (\bibinfo {year}
  {2013})}\BibitemShut {NoStop}%
\bibitem [{\citenamefont {Pla}\ \emph {et~al.}(2012)\citenamefont {Pla},
  \citenamefont {Tan}, \citenamefont {Dehollain}, \citenamefont {Lim},
  \citenamefont {Morton}, \citenamefont {Jamieson}, \citenamefont {Dzurak},\
  and\ \citenamefont {Morello}}]{pla12}%
  \BibitemOpen
  \bibfield  {author} {\bibinfo {author} {\bibfnamefont {J.~J.}\ \bibnamefont
  {Pla}}, \bibinfo {author} {\bibfnamefont {K.~Y.}\ \bibnamefont {Tan}},
  \bibinfo {author} {\bibfnamefont {J.~P.}\ \bibnamefont {Dehollain}}, \bibinfo
  {author} {\bibfnamefont {W.~H.}\ \bibnamefont {Lim}}, \bibinfo {author}
  {\bibfnamefont {J.~J.~L.}\ \bibnamefont {Morton}}, \bibinfo {author}
  {\bibfnamefont {D.~N.}\ \bibnamefont {Jamieson}}, \bibinfo {author}
  {\bibfnamefont {A.~S.}\ \bibnamefont {Dzurak}}, \ and\ \bibinfo {author}
  {\bibfnamefont {A.}~\bibnamefont {Morello}},\ }\href {\doibase
  10.1038/nature11449} {\bibfield  {journal} {\bibinfo  {journal} {Nature}\
  }\textbf {\bibinfo {volume} {489}},\ \bibinfo {pages} {541} (\bibinfo {year}
  {2012})}\BibitemShut {NoStop}%
\bibitem [{\citenamefont {Widmann}\ \emph {et~al.}(2015)\citenamefont
  {Widmann}, \citenamefont {Lee}, \citenamefont {Rendler}, \citenamefont {Son},
  \citenamefont {Fedder}, \citenamefont {Paik}, \citenamefont {Yang},
  \citenamefont {Zhao}, \citenamefont {Yang}, \citenamefont {Booker},
  \citenamefont {Denisenko}, \citenamefont {Jamali}, \citenamefont
  {Momenzadeh}, \citenamefont {Gerhardt}, \citenamefont {Ohshima},
  \citenamefont {Gali}, \citenamefont {Janzen},\ and\ \citenamefont
  {Wrachtrup}}]{widmann15}%
  \BibitemOpen
  \bibfield  {author} {\bibinfo {author} {\bibfnamefont {M.}~\bibnamefont
  {Widmann}}, \bibinfo {author} {\bibfnamefont {S.}~\bibnamefont {Lee}},
  \bibinfo {author} {\bibfnamefont {T.}~\bibnamefont {Rendler}}, \bibinfo
  {author} {\bibfnamefont {N.~T.}\ \bibnamefont {Son}}, \bibinfo {author}
  {\bibfnamefont {H.}~\bibnamefont {Fedder}}, \bibinfo {author} {\bibfnamefont
  {S.}~\bibnamefont {Paik}}, \bibinfo {author} {\bibfnamefont {L.}~\bibnamefont
  {Yang}}, \bibinfo {author} {\bibfnamefont {N.}~\bibnamefont {Zhao}}, \bibinfo
  {author} {\bibfnamefont {S.}~\bibnamefont {Yang}}, \bibinfo {author}
  {\bibfnamefont {I.}~\bibnamefont {Booker}}, \bibinfo {author} {\bibfnamefont
  {A.}~\bibnamefont {Denisenko}}, \bibinfo {author} {\bibfnamefont
  {M.}~\bibnamefont {Jamali}}, \bibinfo {author} {\bibfnamefont {S.~A.}\
  \bibnamefont {Momenzadeh}}, \bibinfo {author} {\bibfnamefont
  {I.}~\bibnamefont {Gerhardt}}, \bibinfo {author} {\bibfnamefont
  {T.}~\bibnamefont {Ohshima}}, \bibinfo {author} {\bibfnamefont
  {A.}~\bibnamefont {Gali}}, \bibinfo {author} {\bibfnamefont {E.}~\bibnamefont
  {Janzen}}, \ and\ \bibinfo {author} {\bibfnamefont {J.}~\bibnamefont
  {Wrachtrup}},\ }\href {\doibase 10.1038/nmat4145} {\bibfield  {journal}
  {\bibinfo  {journal} {Nature Materials}\ }\textbf {\bibinfo {volume} {14}},\
  \bibinfo {pages} {164} (\bibinfo {year} {2015})}\BibitemShut {NoStop}%
\bibitem [{\citenamefont {Kolesov}\ \emph {et~al.}(2012)\citenamefont
  {Kolesov}, \citenamefont {Xia}, \citenamefont {Reuter}, \citenamefont
  {Stohr}, \citenamefont {Zappe}, \citenamefont {Meijer}, \citenamefont
  {Hemmer},\ and\ \citenamefont {Wrachtrup}}]{kolesov12}%
  \BibitemOpen
  \bibfield  {author} {\bibinfo {author} {\bibfnamefont {R.}~\bibnamefont
  {Kolesov}}, \bibinfo {author} {\bibfnamefont {K.}~\bibnamefont {Xia}},
  \bibinfo {author} {\bibfnamefont {R.}~\bibnamefont {Reuter}}, \bibinfo
  {author} {\bibfnamefont {R.}~\bibnamefont {Stohr}}, \bibinfo {author}
  {\bibfnamefont {A.}~\bibnamefont {Zappe}}, \bibinfo {author} {\bibfnamefont
  {J.}~\bibnamefont {Meijer}}, \bibinfo {author} {\bibfnamefont {P.~R.}\
  \bibnamefont {Hemmer}}, \ and\ \bibinfo {author} {\bibfnamefont
  {J.}~\bibnamefont {Wrachtrup}},\ }\href {\doibase 10.1038/ncomms2034}
  {\bibfield  {journal} {\bibinfo  {journal} {Nature Communications}\ }\textbf
  {\bibinfo {volume} {3}},\ \bibinfo {pages} {1029} (\bibinfo {year}
  {2012})}\BibitemShut {NoStop}%
\bibitem [{\citenamefont {Balasubramanian}\ \emph {et~al.}(2008)\citenamefont
  {Balasubramanian}, \citenamefont {Chan}, \citenamefont {Kolesov},
  \citenamefont {Al-Hmoud}, \citenamefont {Tisler}, \citenamefont {Shin},
  \citenamefont {Kim}, \citenamefont {Wojcik}, \citenamefont {Hemmer},
  \citenamefont {Krueger}, \citenamefont {Hanke}, \citenamefont
  {Leitenstorfer}, \citenamefont {Bratschitsch}, \citenamefont {Jelezko},\ and\
  \citenamefont {Wrachtrup}}]{balasubramanian08}%
  \BibitemOpen
  \bibfield  {author} {\bibinfo {author} {\bibfnamefont {G.}~\bibnamefont
  {Balasubramanian}}, \bibinfo {author} {\bibfnamefont {I.~Y.}\ \bibnamefont
  {Chan}}, \bibinfo {author} {\bibfnamefont {R.}~\bibnamefont {Kolesov}},
  \bibinfo {author} {\bibfnamefont {M.}~\bibnamefont {Al-Hmoud}}, \bibinfo
  {author} {\bibfnamefont {J.}~\bibnamefont {Tisler}}, \bibinfo {author}
  {\bibfnamefont {C.}~\bibnamefont {Shin}}, \bibinfo {author} {\bibfnamefont
  {C.}~\bibnamefont {Kim}}, \bibinfo {author} {\bibfnamefont {A.}~\bibnamefont
  {Wojcik}}, \bibinfo {author} {\bibfnamefont {P.~R.}\ \bibnamefont {Hemmer}},
  \bibinfo {author} {\bibfnamefont {A.}~\bibnamefont {Krueger}}, \bibinfo
  {author} {\bibfnamefont {T.}~\bibnamefont {Hanke}}, \bibinfo {author}
  {\bibfnamefont {A.}~\bibnamefont {Leitenstorfer}}, \bibinfo {author}
  {\bibfnamefont {R.}~\bibnamefont {Bratschitsch}}, \bibinfo {author}
  {\bibfnamefont {F.}~\bibnamefont {Jelezko}}, \ and\ \bibinfo {author}
  {\bibfnamefont {J.}~\bibnamefont {Wrachtrup}},\ }\href {\doibase
  10.1038/nature07278} {\bibfield  {journal} {\bibinfo  {journal} {Nature}\
  }\textbf {\bibinfo {volume} {455}},\ \bibinfo {eid} {648} (\bibinfo {year}
  {2008})}\BibitemShut {NoStop}%
\bibitem [{\citenamefont {Maletinsky}\ \emph {et~al.}(2012)\citenamefont
  {Maletinsky}, \citenamefont {Hong}, \citenamefont {Grinolds}, \citenamefont
  {Hausmann}, \citenamefont {Lukin}, \citenamefont {Walsworth}, \citenamefont
  {Loncar},\ and\ \citenamefont {Yacoby}}]{maletinsky12}%
  \BibitemOpen
  \bibfield  {author} {\bibinfo {author} {\bibfnamefont {P.}~\bibnamefont
  {Maletinsky}}, \bibinfo {author} {\bibfnamefont {S.}~\bibnamefont {Hong}},
  \bibinfo {author} {\bibfnamefont {M.~S.}\ \bibnamefont {Grinolds}}, \bibinfo
  {author} {\bibfnamefont {B.}~\bibnamefont {Hausmann}}, \bibinfo {author}
  {\bibfnamefont {M.~D.}\ \bibnamefont {Lukin}}, \bibinfo {author}
  {\bibfnamefont {R.~L.}\ \bibnamefont {Walsworth}}, \bibinfo {author}
  {\bibfnamefont {M.}~\bibnamefont {Loncar}}, \ and\ \bibinfo {author}
  {\bibfnamefont {A.}~\bibnamefont {Yacoby}},\ }\href {\doibase
  10.1038/NNANO.2012.50} {\bibfield  {journal} {\bibinfo  {journal} {Nat.
  Nanotechnol.}\ }\textbf {\bibinfo {volume} {7}},\ \bibinfo {pages} {320}
  (\bibinfo {year} {2012})}\BibitemShut {NoStop}%
\bibitem [{\citenamefont {Cywinski}\ \emph {et~al.}(2008)\citenamefont
  {Cywinski}, \citenamefont {Lutchyn}, \citenamefont {Nave},\ and\
  \citenamefont {Sarma}}]{cywinski08}%
  \BibitemOpen
  \bibfield  {author} {\bibinfo {author} {\bibfnamefont {L.}~\bibnamefont
  {Cywinski}}, \bibinfo {author} {\bibfnamefont {R.~M.}\ \bibnamefont
  {Lutchyn}}, \bibinfo {author} {\bibfnamefont {C.~P.}\ \bibnamefont {Nave}}, \
  and\ \bibinfo {author} {\bibfnamefont {S.~D.}\ \bibnamefont {Sarma}},\ }\href
  {\doibase 10.1103/PhysRevB.77.174509} {\bibfield  {journal} {\bibinfo
  {journal} {Phys. Rev. B}\ }\textbf {\bibinfo {volume} {77}},\ \bibinfo
  {pages} {174509} (\bibinfo {year} {2008})}\BibitemShut {NoStop}%
\bibitem [{\citenamefont {Kotler}\ \emph {et~al.}(2011)\citenamefont {Kotler},
  \citenamefont {Akerman}, \citenamefont {Glickman}, \citenamefont {Keselman},\
  and\ \citenamefont {Ozeri}}]{kotler11}%
  \BibitemOpen
  \bibfield  {author} {\bibinfo {author} {\bibfnamefont {S.}~\bibnamefont
  {Kotler}}, \bibinfo {author} {\bibfnamefont {N.}~\bibnamefont {Akerman}},
  \bibinfo {author} {\bibfnamefont {Y.}~\bibnamefont {Glickman}}, \bibinfo
  {author} {\bibfnamefont {A.}~\bibnamefont {Keselman}}, \ and\ \bibinfo
  {author} {\bibfnamefont {R.}~\bibnamefont {Ozeri}},\ }\href {\doibase
  10.1038/nature10010} {\bibfield  {journal} {\bibinfo  {journal} {Nature}\
  }\textbf {\bibinfo {volume} {473}},\ \bibinfo {pages} {61} (\bibinfo {year}
  {2011})}\BibitemShut {NoStop}%
\bibitem [{\citenamefont {Lange}\ \emph {et~al.}(2011)\citenamefont {Lange},
  \citenamefont {Riste}, \citenamefont {Dobrovitski},\ and\ \citenamefont
  {Hanson}}]{delange11}%
  \BibitemOpen
  \bibfield  {author} {\bibinfo {author} {\bibfnamefont {G.~D.}\ \bibnamefont
  {Lange}}, \bibinfo {author} {\bibfnamefont {D.}~\bibnamefont {Riste}},
  \bibinfo {author} {\bibfnamefont {V.~V.}\ \bibnamefont {Dobrovitski}}, \ and\
  \bibinfo {author} {\bibfnamefont {R.}~\bibnamefont {Hanson}},\ }\href
  {\doibase 10.1103/PhysRevLett.106.080802} {\bibfield  {journal} {\bibinfo
  {journal} {Phys. Rev. Lett.}\ }\textbf {\bibinfo {volume} {106}},\ \bibinfo
  {pages} {080802} (\bibinfo {year} {2011})}\BibitemShut {NoStop}%
\bibitem [{\citenamefont {Alvarez}\ and\ \citenamefont
  {Suter}(2011)}]{alvarez11}%
  \BibitemOpen
  \bibfield  {author} {\bibinfo {author} {\bibfnamefont {G.~A.}\ \bibnamefont
  {Alvarez}}\ and\ \bibinfo {author} {\bibfnamefont {D.}~\bibnamefont
  {Suter}},\ }\href {\doibase 10.1103/PhysRevLett.107.230501} {\bibfield
  {journal} {\bibinfo  {journal} {Phys. Rev. Lett.}\ }\textbf {\bibinfo
  {volume} {107}},\ \bibinfo {pages} {230501} (\bibinfo {year}
  {2011})}\BibitemShut {NoStop}%
\bibitem [{\citenamefont {Yan}\ \emph {et~al.}(2013)\citenamefont {Yan},
  \citenamefont {Gustavsson}, \citenamefont {Bylander}, \citenamefont {Jin},
  \citenamefont {Yoshihara}, \citenamefont {Cory}, \citenamefont {Nakamura},
  \citenamefont {Orlando},\ and\ \citenamefont {Oliver}}]{yan13}%
  \BibitemOpen
  \bibfield  {author} {\bibinfo {author} {\bibfnamefont {F.}~\bibnamefont
  {Yan}}, \bibinfo {author} {\bibfnamefont {S.}~\bibnamefont {Gustavsson}},
  \bibinfo {author} {\bibfnamefont {J.}~\bibnamefont {Bylander}}, \bibinfo
  {author} {\bibfnamefont {X.}~\bibnamefont {Jin}}, \bibinfo {author}
  {\bibfnamefont {F.}~\bibnamefont {Yoshihara}}, \bibinfo {author}
  {\bibfnamefont {D.~G.}\ \bibnamefont {Cory}}, \bibinfo {author}
  {\bibfnamefont {Y.}~\bibnamefont {Nakamura}}, \bibinfo {author}
  {\bibfnamefont {T.~P.}\ \bibnamefont {Orlando}}, \ and\ \bibinfo {author}
  {\bibfnamefont {W.~D.}\ \bibnamefont {Oliver}},\ }\href {\doibase
  10.1038/ncomms3337} {\bibfield  {journal} {\bibinfo  {journal} {Nat. Comms.}\
  }\textbf {\bibinfo {volume} {4}},\ \bibinfo {pages} {2337} (\bibinfo {year}
  {2013})}\BibitemShut {NoStop}%
\bibitem [{\citenamefont {Loretz}\ \emph {et~al.}(2013)\citenamefont {Loretz},
  \citenamefont {Rosskopf},\ and\ \citenamefont {Degen}}]{loretz13}%
  \BibitemOpen
  \bibfield  {author} {\bibinfo {author} {\bibfnamefont {M.}~\bibnamefont
  {Loretz}}, \bibinfo {author} {\bibfnamefont {T.}~\bibnamefont {Rosskopf}}, \
  and\ \bibinfo {author} {\bibfnamefont {C.~L.}\ \bibnamefont {Degen}},\ }\href
  {\doibase 10.1103/PhysRevLett.110.017602} {\bibfield  {journal} {\bibinfo
  {journal} {Phys. Rev. Lett.}\ }\textbf {\bibinfo {volume} {110}},\ \bibinfo
  {pages} {017602} (\bibinfo {year} {2013})}\BibitemShut {NoStop}%
\bibitem [{\citenamefont {Lang}\ \emph {et~al.}(2015)\citenamefont {Lang},
  \citenamefont {Liu},\ and\ \citenamefont {Monteiro}}]{lang15}%
  \BibitemOpen
  \bibfield  {author} {\bibinfo {author} {\bibfnamefont {J.~E.}\ \bibnamefont
  {Lang}}, \bibinfo {author} {\bibfnamefont {R.~B.}\ \bibnamefont {Liu}}, \
  and\ \bibinfo {author} {\bibfnamefont {T.~S.}\ \bibnamefont {Monteiro}},\
  }\href {\doibase 10.1103/PhysRevX.5.041016} {\bibfield  {journal} {\bibinfo
  {journal} {Phys. Rev. X}\ }\textbf {\bibinfo {volume} {5}},\ \bibinfo {pages}
  {041016} (\bibinfo {year} {2015})}\BibitemShut {NoStop}%
\bibitem [{\citenamefont {Laraoui}\ \emph {et~al.}(2010)\citenamefont
  {Laraoui}, \citenamefont {Hodges},\ and\ \citenamefont
  {Meriles}}]{laraoui10}%
  \BibitemOpen
  \bibfield  {author} {\bibinfo {author} {\bibfnamefont {A.}~\bibnamefont
  {Laraoui}}, \bibinfo {author} {\bibfnamefont {J.~S.}\ \bibnamefont {Hodges}},
  \ and\ \bibinfo {author} {\bibfnamefont {C.~A.}\ \bibnamefont {Meriles}},\
  }\href {\doibase 10.1063/1.3497004} {\bibfield  {journal} {\bibinfo
  {journal} {Applied Physics Letters}\ }\textbf {\bibinfo {volume} {97}},\
  \bibinfo {pages} {143104} (\bibinfo {year} {2010})}\BibitemShut {NoStop}%
\bibitem [{\citenamefont {Laraoui}\ \emph {et~al.}(2013)\citenamefont
  {Laraoui}, \citenamefont {Dolde}, \citenamefont {Burk}, \citenamefont
  {Reinhard}, \citenamefont {Wrachtrup},\ and\ \citenamefont
  {Meriles}}]{laraoui13}%
  \BibitemOpen
  \bibfield  {author} {\bibinfo {author} {\bibfnamefont {A.}~\bibnamefont
  {Laraoui}}, \bibinfo {author} {\bibfnamefont {F.}~\bibnamefont {Dolde}},
  \bibinfo {author} {\bibfnamefont {C.}~\bibnamefont {Burk}}, \bibinfo {author}
  {\bibfnamefont {F.}~\bibnamefont {Reinhard}}, \bibinfo {author}
  {\bibfnamefont {J.}~\bibnamefont {Wrachtrup}}, \ and\ \bibinfo {author}
  {\bibfnamefont {C.~A.}\ \bibnamefont {Meriles}},\ }\href {\doibase
  10.1038/ncomms2685} {\bibfield  {journal} {\bibinfo  {journal} {Nature
  Commun.}\ }\textbf {\bibinfo {volume} {4}},\ \bibinfo {pages} {1651}
  (\bibinfo {year} {2013})}\BibitemShut {NoStop}%
\bibitem [{\citenamefont {Zaiser}\ \emph {et~al.}(2016)\citenamefont {Zaiser},
  \citenamefont {Rendler}, \citenamefont {Jakobi}, \citenamefont {Wolf},
  \citenamefont {Lee}, \citenamefont {Wagner}, \citenamefont {Bergholm},
  \citenamefont {Schulte-herbruggen}, \citenamefont {Neumann},\ and\
  \citenamefont {Wrachtrup}}]{zaiser16}%
  \BibitemOpen
  \bibfield  {author} {\bibinfo {author} {\bibfnamefont {S.}~\bibnamefont
  {Zaiser}}, \bibinfo {author} {\bibfnamefont {T.}~\bibnamefont {Rendler}},
  \bibinfo {author} {\bibfnamefont {I.}~\bibnamefont {Jakobi}}, \bibinfo
  {author} {\bibfnamefont {T.}~\bibnamefont {Wolf}}, \bibinfo {author}
  {\bibfnamefont {S.}~\bibnamefont {Lee}}, \bibinfo {author} {\bibfnamefont
  {S.}~\bibnamefont {Wagner}}, \bibinfo {author} {\bibfnamefont
  {V.}~\bibnamefont {Bergholm}}, \bibinfo {author} {\bibfnamefont
  {T.}~\bibnamefont {Schulte-herbruggen}}, \bibinfo {author} {\bibfnamefont
  {P.}~\bibnamefont {Neumann}}, \ and\ \bibinfo {author} {\bibfnamefont
  {J.}~\bibnamefont {Wrachtrup}},\ }\href {\doibase 10.1038/ncomms12279}
  {\bibfield  {journal} {\bibinfo  {journal} {Nature Communications}\ }\textbf
  {\bibinfo {volume} {7}},\ \bibinfo {pages} {12279} (\bibinfo {year}
  {2016})}\BibitemShut {NoStop}%
\bibitem [{\citenamefont {Rosskopf}\ \emph {et~al.}(2016)\citenamefont
  {Rosskopf}, \citenamefont {Zopes}, \citenamefont {Boss},\ and\ \citenamefont
  {Degen}}]{rosskopf16}%
  \BibitemOpen
  \bibfield  {author} {\bibinfo {author} {\bibfnamefont {T.}~\bibnamefont
  {Rosskopf}}, \bibinfo {author} {\bibfnamefont {J.}~\bibnamefont {Zopes}},
  \bibinfo {author} {\bibfnamefont {J.~M.}\ \bibnamefont {Boss}}, \ and\
  \bibinfo {author} {\bibfnamefont {C.~L.}\ \bibnamefont {Degen}},\ }\href
  {http://arxiv.org/abs/1610.03253} {\bibfield  {journal} {\bibinfo  {journal}
  {arXiv:1610.03253}\ } (\bibinfo {year} {2016})}\BibitemShut {NoStop}%
\bibitem [{\citenamefont {Pfender}\ \emph {et~al.}(2016)\citenamefont
  {Pfender}, \citenamefont {Aslam}, \citenamefont {Sumiya}, \citenamefont
  {Onoda}, \citenamefont {Neumann}, \citenamefont {Isoya}, \citenamefont
  {Meriles},\ and\ \citenamefont {Wrachtrup}}]{pfender16}%
  \BibitemOpen
  \bibfield  {author} {\bibinfo {author} {\bibfnamefont {M.}~\bibnamefont
  {Pfender}}, \bibinfo {author} {\bibfnamefont {N.}~\bibnamefont {Aslam}},
  \bibinfo {author} {\bibfnamefont {H.}~\bibnamefont {Sumiya}}, \bibinfo
  {author} {\bibfnamefont {S.}~\bibnamefont {Onoda}}, \bibinfo {author}
  {\bibfnamefont {P.}~\bibnamefont {Neumann}}, \bibinfo {author} {\bibfnamefont
  {J.}~\bibnamefont {Isoya}}, \bibinfo {author} {\bibfnamefont
  {C.}~\bibnamefont {Meriles}}, \ and\ \bibinfo {author} {\bibfnamefont
  {J.}~\bibnamefont {Wrachtrup}},\ }\href {https://arxiv.org/abs/1610.05675}
  {\bibfield  {journal} {\bibinfo  {journal} {arXiv:1610.05675}\ } (\bibinfo
  {year} {2016})}\BibitemShut {NoStop}%
\bibitem [{\citenamefont {Donoho}(2006)}]{donoho06}%
  \BibitemOpen
  \bibfield  {author} {\bibinfo {author} {\bibfnamefont {D.~L.}\ \bibnamefont
  {Donoho}},\ }\href {\doibase 10.1109/TIT.2006.871582} {\bibfield  {journal}
  {\bibinfo  {journal} {IEEE Transactions on Information Theory}\ }\textbf
  {\bibinfo {volume} {52}},\ \bibinfo {pages} {1289} (\bibinfo {year}
  {2006})}\BibitemShut {NoStop}%
\bibitem [{\citenamefont {Rondin}\ \emph {et~al.}(2014)\citenamefont {Rondin},
  \citenamefont {Tetienne}, \citenamefont {Hingant}, \citenamefont {Roch},
  \citenamefont {Maletinsky},\ and\ \citenamefont {Jacques}}]{rondin14}%
  \BibitemOpen
  \bibfield  {author} {\bibinfo {author} {\bibfnamefont {L.}~\bibnamefont
  {Rondin}}, \bibinfo {author} {\bibfnamefont {J.~P.}\ \bibnamefont
  {Tetienne}}, \bibinfo {author} {\bibfnamefont {T.}~\bibnamefont {Hingant}},
  \bibinfo {author} {\bibfnamefont {J.~F.}\ \bibnamefont {Roch}}, \bibinfo
  {author} {\bibfnamefont {P.}~\bibnamefont {Maletinsky}}, \ and\ \bibinfo
  {author} {\bibfnamefont {V.}~\bibnamefont {Jacques}},\ }\href {\doibase
  10.1088/0034-4885/77/5/056503} {\bibfield  {journal} {\bibinfo  {journal}
  {Rep. Prog. Phys.}\ }\textbf {\bibinfo {volume} {77}},\ \bibinfo {pages}
  {056503} (\bibinfo {year} {2014})}\BibitemShut {NoStop}%
\bibitem [{\citenamefont {Schirhagl}\ \emph {et~al.}(2014)\citenamefont
  {Schirhagl}, \citenamefont {Chang}, \citenamefont {Loretz},\ and\
  \citenamefont {Degen}}]{schirhagl14}%
  \BibitemOpen
  \bibfield  {author} {\bibinfo {author} {\bibfnamefont {R.}~\bibnamefont
  {Schirhagl}}, \bibinfo {author} {\bibfnamefont {K.}~\bibnamefont {Chang}},
  \bibinfo {author} {\bibfnamefont {M.}~\bibnamefont {Loretz}}, \ and\ \bibinfo
  {author} {\bibfnamefont {C.~L.}\ \bibnamefont {Degen}},\ }\href {\doibase
  10.1146/annurev-physchem-040513-103659} {\bibfield  {journal} {\bibinfo
  {journal} {Annu. Rev. Phys. Chem.}\ }\textbf {\bibinfo {volume} {65}},\
  \bibinfo {pages} {83} (\bibinfo {year} {2014})}\BibitemShut {NoStop}%
\bibitem [{\citenamefont {Kotler}\ \emph {et~al.}(2013)\citenamefont {Kotler},
  \citenamefont {Akerman}, \citenamefont {Glickman},\ and\ \citenamefont
  {Ozeri}}]{kotler13}%
  \BibitemOpen
  \bibfield  {author} {\bibinfo {author} {\bibfnamefont {S.}~\bibnamefont
  {Kotler}}, \bibinfo {author} {\bibfnamefont {N.}~\bibnamefont {Akerman}},
  \bibinfo {author} {\bibfnamefont {Y.}~\bibnamefont {Glickman}}, \ and\
  \bibinfo {author} {\bibfnamefont {R.}~\bibnamefont {Ozeri}},\ }\href
  {\doibase 10.1103/PhysRevLett.110.110503} {\bibfield  {journal} {\bibinfo
  {journal} {Phys. Rev. Lett.}\ }\textbf {\bibinfo {volume} {110}},\ \bibinfo
  {pages} {110503} (\bibinfo {year} {2013})}\BibitemShut {NoStop}%
\bibitem [{\citenamefont {Jiang}\ \emph {et~al.}(2009)\citenamefont {Jiang},
  \citenamefont {Hodges}, \citenamefont {Maze}, \citenamefont {Maurer},
  \citenamefont {Taylor}, \citenamefont {Cory}, \citenamefont {Hemmer},
  \citenamefont {Walsworth}, \citenamefont {Yacoby}, \citenamefont {Zibrov},\
  and\ \citenamefont {Lukin}}]{jiang09}%
  \BibitemOpen
  \bibfield  {author} {\bibinfo {author} {\bibfnamefont {L.}~\bibnamefont
  {Jiang}}, \bibinfo {author} {\bibfnamefont {J.~S.}\ \bibnamefont {Hodges}},
  \bibinfo {author} {\bibfnamefont {J.~R.}\ \bibnamefont {Maze}}, \bibinfo
  {author} {\bibfnamefont {P.}~\bibnamefont {Maurer}}, \bibinfo {author}
  {\bibfnamefont {J.~M.}\ \bibnamefont {Taylor}}, \bibinfo {author}
  {\bibfnamefont {D.~G.}\ \bibnamefont {Cory}}, \bibinfo {author}
  {\bibfnamefont {P.~R.}\ \bibnamefont {Hemmer}}, \bibinfo {author}
  {\bibfnamefont {R.~L.}\ \bibnamefont {Walsworth}}, \bibinfo {author}
  {\bibfnamefont {A.}~\bibnamefont {Yacoby}}, \bibinfo {author} {\bibfnamefont
  {A.~S.}\ \bibnamefont {Zibrov}}, \ and\ \bibinfo {author} {\bibfnamefont
  {M.~D.}\ \bibnamefont {Lukin}},\ }\href {\doibase 10.1126/science.1176496}
  {\bibfield  {journal} {\bibinfo  {journal} {Science}\ }\textbf {\bibinfo
  {volume} {326}},\ \bibinfo {pages} {267} (\bibinfo {year}
  {2009})}\BibitemShut {NoStop}%
\bibitem [{\citenamefont {Neumann}\ \emph {et~al.}(2010)\citenamefont
  {Neumann}, \citenamefont {Beck}, \citenamefont {Steiner}, \citenamefont
  {Rempp}, \citenamefont {Fedder}, \citenamefont {Hemmer}, \citenamefont
  {Wrachtrup},\ and\ \citenamefont {Jelezko}}]{neumann10science}%
  \BibitemOpen
  \bibfield  {author} {\bibinfo {author} {\bibfnamefont {P.}~\bibnamefont
  {Neumann}}, \bibinfo {author} {\bibfnamefont {J.}~\bibnamefont {Beck}},
  \bibinfo {author} {\bibfnamefont {M.}~\bibnamefont {Steiner}}, \bibinfo
  {author} {\bibfnamefont {F.}~\bibnamefont {Rempp}}, \bibinfo {author}
  {\bibfnamefont {H.}~\bibnamefont {Fedder}}, \bibinfo {author} {\bibfnamefont
  {P.~R.}\ \bibnamefont {Hemmer}}, \bibinfo {author} {\bibfnamefont
  {J.}~\bibnamefont {Wrachtrup}}, \ and\ \bibinfo {author} {\bibfnamefont
  {F.}~\bibnamefont {Jelezko}},\ }\href {\doibase 10.1126/science.1189075}
  {\bibfield  {journal} {\bibinfo  {journal} {Science}\ }\textbf {\bibinfo
  {volume} {329}},\ \bibinfo {pages} {542} (\bibinfo {year}
  {2010})}\BibitemShut {NoStop}%
\bibitem [{\citenamefont {Candes}\ \emph {et~al.}(2006)\citenamefont {Candes},
  \citenamefont {Romberg},\ and\ \citenamefont {Tao}}]{candes06}%
  \BibitemOpen
  \bibfield  {author} {\bibinfo {author} {\bibfnamefont {E.~J.}\ \bibnamefont
  {Candes}}, \bibinfo {author} {\bibfnamefont {J.}~\bibnamefont {Romberg}}, \
  and\ \bibinfo {author} {\bibfnamefont {T.}~\bibnamefont {Tao}},\ }\href@noop
  {} {\bibfield  {journal} {\bibinfo  {journal} {IEEE Transactions on
  Information Theory}\ }\textbf {\bibinfo {volume} {52}},\ \bibinfo {pages}
  {489} (\bibinfo {year} {2006})}\BibitemShut {NoStop}%
\bibitem [{\citenamefont {Sun}\ \emph {et~al.}(2012)\citenamefont {Sun},
  \citenamefont {Chiu}, \citenamefont {Jiang}, \citenamefont {Nallanathan},\
  and\ \citenamefont {Poor}}]{sun12}%
  \BibitemOpen
  \bibfield  {author} {\bibinfo {author} {\bibfnamefont {H.}~\bibnamefont
  {Sun}}, \bibinfo {author} {\bibfnamefont {W.~Y.}\ \bibnamefont {Chiu}},
  \bibinfo {author} {\bibfnamefont {J.}~\bibnamefont {Jiang}}, \bibinfo
  {author} {\bibfnamefont {A.}~\bibnamefont {Nallanathan}}, \ and\ \bibinfo
  {author} {\bibfnamefont {H.~V.}\ \bibnamefont {Poor}},\ }\href@noop {}
  {\bibfield  {journal} {\bibinfo  {journal} {IEEE Transactions on Signal
  Processing}\ }\textbf {\bibinfo {volume} {60}},\ \bibinfo {pages} {6068}
  (\bibinfo {year} {2012})}\BibitemShut {NoStop}%
\bibitem [{\citenamefont {Mamin}\ \emph {et~al.}(2013)\citenamefont {Mamin},
  \citenamefont {Kim}, \citenamefont {Sherwood}, \citenamefont {Rettner},
  \citenamefont {Ohno}, \citenamefont {Awschalom},\ and\ \citenamefont
  {Rugar}}]{mamin13}%
  \BibitemOpen
  \bibfield  {author} {\bibinfo {author} {\bibfnamefont {H.~J.}\ \bibnamefont
  {Mamin}}, \bibinfo {author} {\bibfnamefont {M.}~\bibnamefont {Kim}}, \bibinfo
  {author} {\bibfnamefont {M.~H.}\ \bibnamefont {Sherwood}}, \bibinfo {author}
  {\bibfnamefont {C.~T.}\ \bibnamefont {Rettner}}, \bibinfo {author}
  {\bibfnamefont {K.}~\bibnamefont {Ohno}}, \bibinfo {author} {\bibfnamefont
  {D.~D.}\ \bibnamefont {Awschalom}}, \ and\ \bibinfo {author} {\bibfnamefont
  {D.}~\bibnamefont {Rugar}},\ }\href {\doibase 10.1126/science.1231540}
  {\bibfield  {journal} {\bibinfo  {journal} {Science}\ }\textbf {\bibinfo
  {volume} {339}},\ \bibinfo {pages} {557} (\bibinfo {year}
  {2013})}\BibitemShut {NoStop}%
\bibitem [{\citenamefont {Staudacher}\ \emph {et~al.}(2013)\citenamefont
  {Staudacher}, \citenamefont {Shi}, \citenamefont {Pezzagna}, \citenamefont
  {Meijer}, \citenamefont {Du}, \citenamefont {Meriles}, \citenamefont
  {Reinhard},\ and\ \citenamefont {Wrachtrup}}]{staudacher13}%
  \BibitemOpen
  \bibfield  {author} {\bibinfo {author} {\bibfnamefont {T.}~\bibnamefont
  {Staudacher}}, \bibinfo {author} {\bibfnamefont {F.}~\bibnamefont {Shi}},
  \bibinfo {author} {\bibfnamefont {S.}~\bibnamefont {Pezzagna}}, \bibinfo
  {author} {\bibfnamefont {J.}~\bibnamefont {Meijer}}, \bibinfo {author}
  {\bibfnamefont {J.}~\bibnamefont {Du}}, \bibinfo {author} {\bibfnamefont
  {C.~A.}\ \bibnamefont {Meriles}}, \bibinfo {author} {\bibfnamefont
  {F.}~\bibnamefont {Reinhard}}, \ and\ \bibinfo {author} {\bibfnamefont
  {J.}~\bibnamefont {Wrachtrup}},\ }\href {\doibase 10.1126/science.1231675}
  {\bibfield  {journal} {\bibinfo  {journal} {Science}\ }\textbf {\bibinfo
  {volume} {339}},\ \bibinfo {pages} {561} (\bibinfo {year}
  {2013})}\BibitemShut {NoStop}%
\bibitem [{\citenamefont {Loretz}\ \emph {et~al.}(2014)\citenamefont {Loretz},
  \citenamefont {Pezzagna}, \citenamefont {Meijer},\ and\ \citenamefont
  {Degen}}]{loretz14apl}%
  \BibitemOpen
  \bibfield  {author} {\bibinfo {author} {\bibfnamefont {M.}~\bibnamefont
  {Loretz}}, \bibinfo {author} {\bibfnamefont {S.}~\bibnamefont {Pezzagna}},
  \bibinfo {author} {\bibfnamefont {J.}~\bibnamefont {Meijer}}, \ and\ \bibinfo
  {author} {\bibfnamefont {C.~L.}\ \bibnamefont {Degen}},\ }\href {\doibase
  10.1063/1.4862749} {\bibfield  {journal} {\bibinfo  {journal} {Appl. Phys.
  Lett.}\ }\textbf {\bibinfo {volume} {104}},\ \bibinfo {pages} {33102}
  (\bibinfo {year} {2014})}\BibitemShut {NoStop}%
\bibitem [{\citenamefont {Cai}\ \emph {et~al.}(2013)\citenamefont {Cai},
  \citenamefont {Retzker}, \citenamefont {Jelezko},\ and\ \citenamefont
  {Plenio}}]{cai13}%
  \BibitemOpen
  \bibfield  {author} {\bibinfo {author} {\bibfnamefont {J.~M.}\ \bibnamefont
  {Cai}}, \bibinfo {author} {\bibfnamefont {A.}~\bibnamefont {Retzker}},
  \bibinfo {author} {\bibfnamefont {F.}~\bibnamefont {Jelezko}}, \ and\
  \bibinfo {author} {\bibfnamefont {M.~B.}\ \bibnamefont {Plenio}},\ }\href
  {\doibase 10.1038/nphys2519} {\bibfield  {journal} {\bibinfo  {journal}
  {Nature Physics}\ }\textbf {\bibinfo {volume} {9}},\ \bibinfo {pages} {168}
  (\bibinfo {year} {2013})}\BibitemShut {NoStop}%
\bibitem [{\citenamefont {Mehring}(2012)}]{mehring12}%
  \BibitemOpen
  \bibfield  {author} {\bibinfo {author} {\bibfnamefont {M.}~\bibnamefont
  {Mehring}},\ }\href
  {http://link.springer.com/book/10.1007%2F978-3-642-68756-3} {\emph {\bibinfo
  {title} {Principles of High Resolution NMR in Solids}}}\ (\bibinfo
  {publisher} {Springer Berlin Heidelberg},\ \bibinfo {year}
  {2012})\BibitemShut {NoStop}%
\bibitem [{\citenamefont {Reif}(2012)}]{reif12}%
  \BibitemOpen
  \bibfield  {author} {\bibinfo {author} {\bibfnamefont {B.}~\bibnamefont
  {Reif}},\ }\href
  {http://www.sciencedirect.com/science/article/pii/S1090780711005969}
  {\bibfield  {journal} {\bibinfo  {journal} {Journal of Magnetic Resonance}\
  }\textbf {\bibinfo {volume} {216}},\ \bibinfo {pages} {1} (\bibinfo {year}
  {2012})}\BibitemShut {NoStop}%
\bibitem [{\citenamefont {Appelt}\ \emph {et~al.}(2006)\citenamefont {Appelt},
  \citenamefont {Kuhn}, \citenamefont {Hasing},\ and\ \citenamefont
  {Blumich}}]{appelt06}%
  \BibitemOpen
  \bibfield  {author} {\bibinfo {author} {\bibfnamefont {S.}~\bibnamefont
  {Appelt}}, \bibinfo {author} {\bibfnamefont {H.}~\bibnamefont {Kuhn}},
  \bibinfo {author} {\bibfnamefont {F.~W.}\ \bibnamefont {Hasing}}, \ and\
  \bibinfo {author} {\bibfnamefont {B.}~\bibnamefont {Blumich}},\ }\href
  {\doibase 10.1038/nphys211} {\bibfield  {journal} {\bibinfo  {journal} {Nat
  Phys}\ }\textbf {\bibinfo {volume} {2}},\ \bibinfo {pages} {105} (\bibinfo
  {year} {2006})}\BibitemShut {NoStop}%
\bibitem [{\citenamefont {Boss}\ \emph {et~al.}(2017)\citenamefont {Boss},
  \citenamefont {Cujia}, \citenamefont {Zopes},\ and\ \citenamefont
  {Degen}}]{boss17}%
  \BibitemOpen
  \bibfield  {author} {\bibinfo {author} {\bibfnamefont {J.~M.}\ \bibnamefont
  {Boss}}, \bibinfo {author} {\bibfnamefont {K.~S.}\ \bibnamefont {Cujia}},
  \bibinfo {author} {\bibfnamefont {J.}~\bibnamefont {Zopes}}, \ and\ \bibinfo
  {author} {\bibfnamefont {C.~L.}\ \bibnamefont {Degen}},\ }\href {\doibase
  10.1126/science.aam7009} {\bibfield  {journal} {\bibinfo  {journal}
  {Science}\ }\textbf {\bibinfo {volume} {356}},\ \bibinfo {pages} {837}
  (\bibinfo {year} {2017})}\BibitemShut {NoStop}%
\bibitem [{\citenamefont {Schmitt}\ \emph {et~al.}(2017)\citenamefont
  {Schmitt}, \citenamefont {Gefen}, \citenamefont {Sturner}, \citenamefont
  {Unden}, \citenamefont {Wolff}, \citenamefont {Muller}, \citenamefont
  {Scheuer}, \citenamefont {Naydenov}, \citenamefont {Markham}, \citenamefont
  {Pezzagna}, \citenamefont {Meijer}, \citenamefont {Schwarz}, \citenamefont
  {Plenio}, \citenamefont {Retzker}, \citenamefont {McGuinness},\ and\
  \citenamefont {Jelezko}}]{schmitt17}%
  \BibitemOpen
  \bibfield  {author} {\bibinfo {author} {\bibfnamefont {S.}~\bibnamefont
  {Schmitt}}, \bibinfo {author} {\bibfnamefont {T.}~\bibnamefont {Gefen}},
  \bibinfo {author} {\bibfnamefont {F.~M.}\ \bibnamefont {Sturner}}, \bibinfo
  {author} {\bibfnamefont {T.}~\bibnamefont {Unden}}, \bibinfo {author}
  {\bibfnamefont {G.}~\bibnamefont {Wolff}}, \bibinfo {author} {\bibfnamefont
  {C.}~\bibnamefont {Muller}}, \bibinfo {author} {\bibfnamefont
  {J.}~\bibnamefont {Scheuer}}, \bibinfo {author} {\bibfnamefont
  {B.}~\bibnamefont {Naydenov}}, \bibinfo {author} {\bibfnamefont
  {M.}~\bibnamefont {Markham}}, \bibinfo {author} {\bibfnamefont
  {S.}~\bibnamefont {Pezzagna}}, \bibinfo {author} {\bibfnamefont
  {J.}~\bibnamefont {Meijer}}, \bibinfo {author} {\bibfnamefont
  {I.}~\bibnamefont {Schwarz}}, \bibinfo {author} {\bibfnamefont
  {M.}~\bibnamefont {Plenio}}, \bibinfo {author} {\bibfnamefont
  {A.}~\bibnamefont {Retzker}}, \bibinfo {author} {\bibfnamefont {L.~P.}\
  \bibnamefont {McGuinness}}, \ and\ \bibinfo {author} {\bibfnamefont
  {F.}~\bibnamefont {Jelezko}},\ }\href {\doibase 10.1126/science.aam5532}
  {\bibfield  {journal} {\bibinfo  {journal} {Science}\ }\textbf {\bibinfo
  {volume} {356}},\ \bibinfo {pages} {832} (\bibinfo {year}
  {2017})}\BibitemShut {NoStop}%
\bibitem [{\citenamefont {Bucher}\ \emph {et~al.}(2017)\citenamefont {Bucher},
  \citenamefont {Glenn}, \citenamefont {Lee}, \citenamefont {Lukin},
  \citenamefont {Park},\ and\ \citenamefont {Walsworth}}]{bucher17}%
  \BibitemOpen
  \bibfield  {author} {\bibinfo {author} {\bibfnamefont {D.~B.}\ \bibnamefont
  {Bucher}}, \bibinfo {author} {\bibfnamefont {D.~R.}\ \bibnamefont {Glenn}},
  \bibinfo {author} {\bibfnamefont {J.}~\bibnamefont {Lee}}, \bibinfo {author}
  {\bibfnamefont {M.~D.}\ \bibnamefont {Lukin}}, \bibinfo {author}
  {\bibfnamefont {H.}~\bibnamefont {Park}}, \ and\ \bibinfo {author}
  {\bibfnamefont {R.~L.}\ \bibnamefont {Walsworth}},\ }\href
  {https://arxiv.org/abs/1705.08887} {\bibfield  {journal} {\bibinfo  {journal}
  {arXiv:1705.08887}\ } (\bibinfo {year} {2017})}\BibitemShut {NoStop}%
\end{thebibliography}
\end{document}

%%%%%%%%%%%%%%%%%%%%%%%%%%%%%%%%%%%%%%%%%%%%%%%%%%%%%%%%%%%%%%%%%%%%%%%%%%%%%%%%%%%%%%%%%%%%%%%%%%%%%%%%%%%%%%%%%%%%%%% title page

\begin{center}
{\LARGE{}Supplementary Materials for} \\
\vspace{0.5cm}
{\Large{} Quantum sensing with arbitrary frequency resolution} \\
\vspace{0.5cm}
J. M. Boss$^\dagger$, K. S. Cujia$^\dagger$, J. Zopes, and C. L. Degen$^\ast$ \\
\vspace{0.5cm}
$^\ast$ correspondence to: degenc@ethz.ch \\
$^\dagger$ these authors contributed equally to this work. \\
\end{center}

\begin{comment}
\vspace{1.0cm}
{\bf This PDF file includes:}
\vspace{0.5cm} \\
Materials and Methods \\
Supplementary Text 1 to 4\\
Figures S1 to S5 \\
%References (31-40)
\end{comment}

%%%%%%%%%%%%%%%%%%%%%%%%%%%%%%%%%%%%%%%%%%%%%%%%%%%%%%%%%

%\newpage{}

\section*{Materials and Methods}

%%%%%
\subsection*{Experimental setup}

Experiments were performed with a custom-built confocal microscope equipped with a green 532~nm excitation laser and a single photon detector, as well as microwave and radio frequency sources to control the NV center spin and the \NN nuclear spin, respectively.  The NV centers were created by $^{15}$N$^+$ ion implantation at an energy of 5~keV and subsequent annealing at 850$^{\circ}$~C.  We chose the \NN species to discriminate implanted NV centers from native (\N) NV centers.  However, the isotope species played no role for the present experiments.  We etched nano-pillars into the diamond surface \cite{babinec10,momenzadeh15} to increase the photon collection efficiency by a factor of 10 to 15 compared to a non-structured diamond surface.  The continuous wave (CW) photon count rate was between 400 and 700~kC/s.

Microwave pulses were synthesized on an arbitary waveform generator (Tektronix AWG5012C) and upconverted to $\sim 10\unit{GHz}$ using a local oscillator (Hittite HMCT2100) and a single-sideband mixer (IQ0618, Marki microwave).  Radio-frequency pulses were synthesized on a second arbitrary waveform generator (NI 5421, National Instruments).  Microwave and radio-frequency pulses were amplified separately and then combined using a bias T.  The pulses were delivered to the NV center using a coplanar waveguide (CPW) deposited on a quartz cover slip in a transmission line geometry.  The transmission line was terminated by an external 50$\,\Omega$ load.

A cylindrical permanent magnet was used to create a magnetic bias field of $450-500\unit{mT}$ at the location of the NV center.  At this high bias field repolarization of the \NN nucleus under optical illumination \cite{aslam13} is greatly suppressed.  This allowed for a large number ($n>1000$) of repetitive nuclear spin readouts to be performed.  The magnetic field direction was aligned with the NV symmetry axis by adjusting the relative location of the permanent magnet.  The alignment was optimized by maximizing the CW photon count rate and by minimizing the depolarization of \NN nuclear spin states under repetitive readout.  The magnetic field drifted by typically a few Gauss over the course of an experiment, corresponding to a variation in the EPR frequency of $\sim 1\unit{MHz}$.  Because the drifts were slow, we could continuously track the EPR resonance during a measurement and adjust the microwave excitation frequency.  In this way, the detuning between the EPR resonance frequency and the microwave frequency could be reduced to $<100\unit{kHz}$.

%%%%%
\subsection*{Sensing sequence}
\label{sec:technical_details_seq}

\begin{figure}[h!]
\centering
\includegraphics[width=0.99\textwidth]{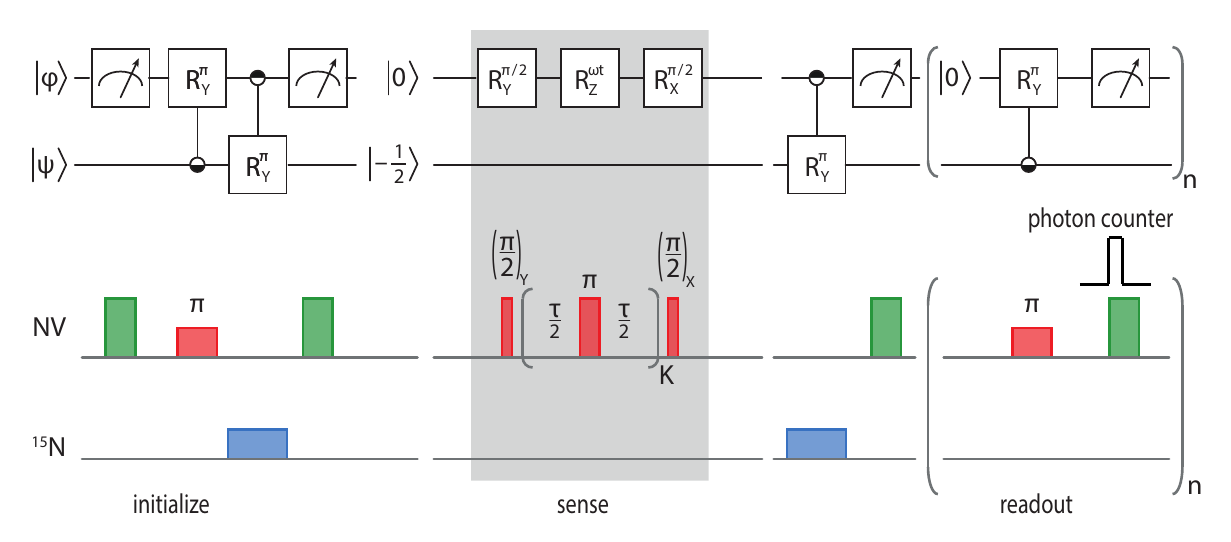}
\caption{Qubit gate diagram of one sensing instance.  Two qubits were used, including a probe qubit (implemented by the electronic spin of the NV center) and a memory qubit (implemented by the \NN nuclear spin of the NV center).  The top channel represents the electronic and the bottom channel the nuclear qubit.
The nuclear spin was initialized by a laser pulse (green) plus two c-NOT gates. Thereafter, the electronic qubit was initialized by another laser pulse.  The c-NOT gates were implemented by selective microwave (red) and radio-frequency (blue) inversion pulses on the electronic and nuclear hyperfine transitions, respectively.  A CPMG sequence adjusted to the frequency of interest was then executed on the electronic qubit. The resulting state was stored in the nuclear qubit via another c-NOT gate and subsequently read out in a repetitive quantum-nondemolition measurement \cite{jiang09}. The readout sequence consisted of the repetitive execution of an electronic c-NOT gate followed by a readout laser pulse of duration $\approx$800~ns.  Full-height pulses symbolize non-selective pulses and half-height pulses symbolize selective pulses.
}
\label{fig:scheme_details}
\end{figure}

A schematic of the sensing sequence is shown in Fig.~\ref{fig:scheme_details}.

To arm the sensor, we initialized both the electronic and the \NN nuclear spins.  The electronic spin was initialized by means of a $\sim 1.5\unit{\us}$ laser pulse.  The nuclear \NN spin was initialized by a sequence of two c-NOT gates followed by a laser pulse to reset the electronic spin.  The initialization efficiency was not measured, but is expected to be $>80\%$ for the electronic spin \cite{jelezko06} and $>70\%$ for the nuclear spin \cite{rosskopf16}, respectively.  The first (electronic) c-NOT gate included a selective microwave $\pi$ pulse on the lower hyperfine resonance ($\sim 9922.22\unit{MHz}$) of the electronic $m_S=0\leftrightarrow m_S=-1$ transition.  The second (nuclear) c-NOT gate included a selective radio-frequency $\pi$ pulse on the higher frequency hyperfine resonance ($\sim 1.97\unit{MHz}$) of the nuclear $m_I=-1/2\leftrightarrow m_I=+1/2$ transition.  The duration of the selective microwave pulse was $\sim 290\unit{ns}$ and the duration of the selective radio-frequency pulse was $\sim 40\unit{\us}$.

Quantum lock-in detection was implemented by a Carr-Purcell-Meiboum-Gibbs (CPMG) sequence of periodic microwave $\pi$ pulses.  The sequence consisted of $K$ pulses with an interpulse delay $\tau$.  The interpulse delay was chosen to approximately match the expected a.c. signal frequency $\fac$ as $\fac \approx m/(2\tau)$, where $m=1,3,5,...$ is the harmonic order of the sequence.  The value of $\tau$ can be determined either by a prior knowledge about $\fac$, or by scanning a range of $\tau$ values.  The total duration of the CPMG sequence was $\ta = K\tau$.  To optimize the sequence, we adjusted the number of pulses $K$ so that the maximum phase pick-up was $\sim 0.5$.

The readout of the final NV state was performed indirectly via a repetitive quantum non-demolition measurement of the \NN nuclear spin \cite{rosskopf16}.  For this purpose, the final electronic spin state was stored in the \NN spin state using a nuclear c-NOT gate and a laser pulse for resetting the electronic spin state.  Next, the nuclear spin state was read out using an electronic c-NOT gate followed by a short ($600 - 800\unit{ns}$) laser pulse.  The nuclear read-out was repeated $n$ times (with $n$ up to 2,000).  The duration of one nuclear readout was $\tr \approx 2.32\unit{\us}$.  The total readout duration was $n\times 2.32\unit{\us}$.  The integrated counts over $n$ repetitive readouts correspond to a single sample record $y_k$.

The sensing sequence incorporated an additional delay time $\td$.  This delay time was used for initialization and for accommodating a separate short pulse sequence to continuously track the NV resonance frequency and correct for drifts.  The delay time was also used to adjust the sampling time $\ts = \ta + \tr + \td$ in the compressed sampling experiment.

%%%%%
\subsection*{Experimental parameters}

The following tables give the parameters that went into the measurements shown in Figures 2, 3, and 4.

\subsubsection*{Parameters for Figure 2A-B}
\begin{table}[h!]
    %\centering
    \begin{tabular}{l|l}
    B field & 546.59$\unit{mT}$\\
    NV initialization laser pulse & $2\unit{\us}$\\
    Repetitive readout laser pulse & $600\unit{ns}$\\
    Selective electronic pulse duration & $290\unit{ns}$\\
    Selective nuclear pulse duration & $30\unit{\us}$\\
    CPMG pulses $K$ & $32$\\
    CPMG duration $\ta$ & $26.622\unit{\us}$\\
    Sampling period $\ts$ & $4.21152 \unit{ms}$\\
    Number of repetitive readouts $n$ & $1000$
    \end{tabular}\hfill\
    \label{tbl:params_fig2}
\end{table}
We have evaluated the amplitude of the a.c. magnetic field detected in this experiment.  The CPMG duration $\ta$ was chosen such that the amplitude of the probability was $\sim 0.25$.  According to Eq. (2), the amplitude of the phase was $\sim 0.5$.  According to Eq. (1), the amplitude of the signal $\Omega$ was $\sim 2\pi\times 4.7\unit{kHz}$.  The amplitude of the a.c. magnetic field was $\Omega/\ye \sim 170\unit{nT}$, where $\ye = 2\pi\times 28\unit{GHz/T}$ is the electron gyromagnetic ratio.

\newpage

\subsubsection*{Parameters for Figures 2C-D and Figure S3}
\begin{table}[h!]
    \centering
    \begin{tabular}{l|l}
    B field & $456.54\unit{mT}$\\
    NV initialization laser pulse & $1\unit{\us}$\\
    Repetitive readout laser pulse & $800\unit{ns}$\\
    Selective electronic pulse duration & $290\unit{ns}$\\
    Selective nuclear pulse duration & $40\unit{\us}$\\
    CPMG pulses $K$ & $16$\\
    CPMG duration $\ta$ & $6.654\unit{\us}$\\
    Sampling period $\ts$ & $1.31524\unit{ms}$\\
    Number of repetitive readouts $n$ & $498$
    \end{tabular}\hfill\
    \label{tbl:params_fig3}
\end{table}

\subsubsection*{Parameters for Figure 3}
\begin{table}[h!]
    \centering
    \begin{tabular}{l|l}
    B field & $456.54\unit{mT}$\\
    NV initialization laser pulse & $1\unit{\us}$\\
    Repetitive readout laser pulse & $800\unit{ns}$\\
    Selective electronic pulse duration & $290\unit{ns}$\\
    Selective nuclear pulse duration & $40\unit{\us}$\\
    CPMG pulses $K$ & $16$\\
    CPMG duration $\ta$ & $6.654\unit{\us}$\\
    Sample records $N$ & $385263$\\
    %Number of repetitive readouts $n$ & up to $2000$
		Sampling period $\ts$ & $1.31524\unit{ms}$ (for datapoints where $n\leq498$) \\
		Sampling period $\ts$ & $5.16468\unit{ms}$ (for datapoints where $n>498$) \\
    \end{tabular}\hfill\
    \label{tbl:params_fig4}
\end{table}
We have used this experiment to determine the optimum SNR for a one-hour measurement interval.  In the plot, the SNR at the threshold gain of $\Cthresh = 27$ ($n=260$) was $1.0\ee{3}$.  The duration of this measurement was $T=N\ts=294\unit{s}$, where $N=3.85\ee{5}$ was the number of samples and $\ts=0.763\unit{ms}$ was the sampling period.  According to Eq. (6), this converts to an SNR of $1.0\ee{3} \times (1\unit{h})/T = 1.2\ee{4}$ for a one-hour interval.

\subsubsection*{Parameters for Figure 4}
\begin{table}[h!]
    \centering
    \begin{tabular}{l|l}
    B field & $456.69\unit{mT}$\\
    NV initialization laser pulse & $1\unit{\us}$\\
    Repetitive readout laser pulse & $800\unit{ns}$\\
    Selective electronic pulse duration & $290\unit{ns}$\\
    Selective nuclear pulse duration & $40\unit{\us}$\\
    CPMG pulses $K$ & $16$\\
    CPMG duration $\ta$ & $19.965\unit{\us}$\\
    Number of repetitive readouts $n$ & $498$
    \end{tabular}\hfill\
    \label{tbl:params_fig5}
\end{table}

\newpage

The following table gives the sampling rates $\fsi$ for the spectra shown in Fig. 4A:
%The following Table gives the sampling rates $\fsi$ and record lengths $N_i$ for the $p=7$ spectra shown in Fig. 5A:
%
\begin{table}[h!]
    \centering
    \begin{tabular}{c|c}
    $i$ & $\fsi$ (Hz) \\
    1 & 752.6719855487 \\
    2 & 750.7507507508 \\
    3 & 749.4454103963 \\
    4 & 747.6747315848 \\
    5 & 746.0904858541 \\
    6 & 743.7819826253 \\
    7 & 742.6772027806
    \end{tabular}\hfill\
    %\caption{Sampling frequencies used for spectrum reconstruction.}
    \label{tbl:unfolding_details}
\end{table}

%%%%%%%%%%%%%%%%%%%%%%%%%%%%%%%%%%%%%%%%%%%%%%%%%%%%%%%%%%%%%%%%%%%%%%%%%%%%%%%%%%%%%%%%%%%%%%%%%%%%%%%%%%%%%

\newpage

\section*{Supplementary Text 1: Details of the quantum lock-in protocol}

\begin{figure}[h!]
\centering
\includegraphics[width=0.99\textwidth]{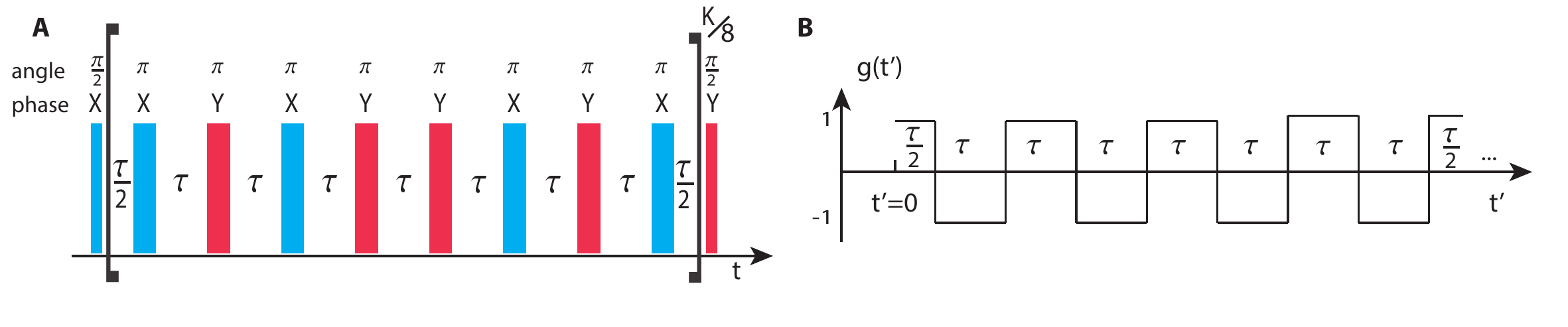}
\caption{
(\textbf{A}) Pulse timing diagram of the CPMG sequence executed on the electronic qubit.
Blue (red) microwave pulses stand for rotations around the $X$-axis ($Y$-axis).  The eight $\pi$-pulses in the square bracket are repeated $\frac{K}{8}$ times.  The alternation of the rotation axes is that of an XY8 sequence \cite{gullion90}.
(\textbf{B}) Modulation function $g(t)$ of the CPMG sequence.  Each $\pi$ reverts the accumulated quantum phase of the qubit, represented by a change in sign of the modulation function.
}
	\label{fig:SuppPulseSeq}
\end{figure}

Our implementation of the quantum lock-in protocol is based on a Carr-Purcell-Meiboum-Gibbs (CPMG)-type sequence of $K$ periodic $\pi$-pulses.  Assuming the qubit is in the $\ket{0}$ state (the $m_S=0$ spin state) at the beginning of the sequence, the first $\left(\pi/2\right)_Y$-pulse rotates it into the $\ket{+X}=\frac{1}{\sqrt2}(\ket{0}+\ket{1})$ state, where $\ket{1}$ corresponds to the $m_S=-1$ spin state.  The qubit then evolves under a series of $\pi$-pulses with inter-pulse delays $\tau$.  At the end of the CPMG sequence, the resulting state is rotated again by $\pi/2$ but this time around the $X$ axis.  This leaves the qubit in the superposition state
\begin{align}
\ket{\psi} \equiv  \sin\left(\frac{\phi}{2}+\frac{\pi}{4}\right) \ket{0}-\cos\left(\frac{\phi}{2}+\frac{\pi}{4}\right) \ket{1} \ ,
\end{align}
where we omit a global phase. $\phi$ is the phase acquired by the qubit during the decoupling sequence.  The readout projects the qubit onto either $\ket{0}$ or $\ket{1}$.  The probability $p$ for projection onto $\ket{1}$ is
\begin{align}
p &= \left|\langle 1|\psi\rangle\right|^2 
   = \frac{1}{2}\left(1 - \sin\phi\right) \label{eq:transitionProb}.
\end{align}
Next we consider how the phase $\phi$ relates to the a.c. signal.  We assume that the a.c. signal is given by the oscillating field
\begin{align}
x(t) = \Omega \cos(2 \pi \fac t +\alpha),
\end{align}
where $\Omega$ is the amplitude (in units of angular frequency), $\fac$ the frequency, and $\alpha$ the initial phase of the signal at time $t=0$ after the first $\pi/2$ pulse was applied.  For the specific situation of our experiment, where the qubit is an electronic spin, the a.c. signal is generated by a magnetic field,
\begin{align}
x(t) = \ye B_z \cos(2 \pi \fac t +\alpha) \ ,
\end{align}
where $B_z=\Omega/\ye$ represents the magnetic field component along the qubit's quantization axis, and $\ye = 2\pi\times 28\unit{GHz/T}$ is the electron gyromagnetic ratio.
For simplicity, we in the following assume that $\alpha=0$ because we always measure a relative time.  The phase acquired by the qubit is then given by
\begin{align}
\phi(t) = \int_0^{\ta} dt' x(t+t') g(t') \ ,
\end{align}
where $g(t') = (-1)^{[t'/\tau]}$ is the modulation function \cite{degen16} of the CPMG sequence (see Fig. \ref{fig:SuppPulseSeq}).  For an even number of pulses $K$, the phase accumulated under the modulation function of the CPMG sequence is \cite{degen16}
\begin{align}
\phi(t) = -\frac{8}{2\pi\fac}\Omega
\cos\left(2\pi \fac \left[t+\frac12K\tau\right]\right)
\frac{\sin\left(2\pi \fac \frac{K\tau}{2}\right)}{\sin\left(2\pi \fac \tau\right)}
\sin\left(2\pi \fac \frac{\tau}{4}\right)^2
\sin\left(2\pi \fac \frac{\tau}{2}\right) \ .
\label{eq:phase}
\end{align}
When the inter-pulse delay $\tau$ is approximately adjusted to the frequency of the a.c. signal, $\tau\approx m/(2\fac)$ (where $m=1,3,5,...$ is the harmonic order), the above general formula simplifies to
\begin{align}
\phi(t)
  &= \left( -1 \right)^q \frac{2\ta}{m\pi} \Omega\cos(2\pi\fac t) 
   =\left( -1 \right)^q \frac{2\ta}{m\pi}x(t),
\end{align}
where $q = \frac{m-1}{2}$. As a result, the phase $\phi(t)$ is directly proportional to the instantaneous value of the signal $x(t)$.  Thus, by using a series of quantum lock-in measurements, we can record how the ac signal $x(t)$ evolves with time.

To calculate the transition probability, we insert $\phi(t)$ into Eq.~\ref{eq:transitionProb},
\begin{align}
p(t)
  &= \frac12\left(1 - \sin\left[ (-1)^q \frac{2\ta}{m\pi}x(t)\right] \right)
   = \frac12\left(1 - \sin\left[ \phimax\cos(2\pi\fac t) \right]\right) \ .
\end{align}
$\phimax$ is the amplitude of the a.c. signal expressed in units of the accumulated phase,
\begin{align}
\phimax &= (-1)^q \frac{2\ta\Omega}{m\pi} \overset{\quad m=1 \quad}{=} \frac{2\ta\Omega}{\pi} \ ,
\label{eq:phimax}
\end{align}
where the last expression ($m=1$) represents our experimental situation.

For small $\phi$, the sine term is linear in $\phi$ and the probability is
\begin{align}
p(t) \approx \frac12\left(1 - \phimax\cos(2\pi\fac t) \right) \ ,\label{eq:linearTransProb}
\end{align}
When several signals are present, the probability $p(t)$ simply is a linear combination of the individual contributions, as long as the maximum phase $\phi$ is within the linear range of the sine.

Conversely, when $\phimax\gtrsim 1$, the response of $p$ becomes nonlinear \cite{kotler13}.  Specifically, for a single signal with frequency $\fac$ and amplitude $\phimax$,
\begin{align}
p(t)
%&= \frac{1}{2}\left\{1 - \sin\left[A\cos\left(2\pi f_{ac} t_k\right)\right]\right\}
  &= \frac12 - \sum^{\infty}_{k=0}(-1)^kJ_{2k+1}(\phimax)\cos\left[(2k+1)2\pi f_{ac} t\right] \ , \label{eq:nonLinearSig}
\end{align}
where $J_k(\phimax)$ is the Bessel function of first kind.  The probability $p(t)$ now contains harmonics at $3\fac$, $5\fc$, etc. of the original signal frequency $\fac$ whose amplitudes are given by Bessel functions.  Fig.~\ref{fig:besselPlots}A shows simulated spectra for values of $\phimax$ between 0.5 and 14.  Fig.~\ref{fig:besselPlots}C further shows that the combined power of all harmonic peaks, given through $\sum^{\infty}_{k=0} J_{2k+1}^2(\phimax)$, saturates as $\phimax\gtrsim 1$ and approaches $0.25$ as $\phimax\rightarrow\infty$.

Finally, if several signals are present with $\phimax$ in the nonlinear regime, frequency mixing occurs (Fig.~\ref{fig:besselPlots}B).
Because of the harmonic generation and frequency mixing, spectra acquired in the nonlinear regime are difficult to interpret and it is advantageous to stay in the linear range of the sensor.

To confirm the theoretical analysis above, we recorded spectra for different values of $\phimax$ exceeding the linear regime. Fig.~\ref{fig:harmonics}A shows these measurements. We assigned the peaks in the spectra to their corresponding harmonic order. Thereby, we find harmonics up to the order $2k+1 = 21$ for the measurement of the strongest signal where $\phimax = 21.6$. Furthermore, in Fig.~\ref{fig:harmonics}B, we fitted the the peak height for the first 4 harmonics to their Bessel functions squared, finding good agreement with Eq.~(\ref{eq:nonLinearSig}).

%%%%%
\newpage
\section*{Supplementary Text 2: Scaling of frequency estimation}
\label{sec:resolutionScaling}

In Fig. 2E of the main text we investigate the scaling of the uncertainty of the estimated center frequency with increasing total measurement time $T$.  We find that the uncertainty scales as $T^{-1.5}$ if the intrinsic linewidth parameter $\Df$ of the signal is smaller than the frequency resolution $\df=1/T$, and that it scales as $T^{-0.5}$ if $\Df$ is larger than $\df$.  This section serves to motivate these two scaling laws.

We first consider the situation where the intrinsic linewidth of the spectral peak is larger than the frequency resolution, $\Df>\df$.  This situation leads to a $T^{-0.5}$ scaling for the uncertainty in the center frequency.  We estimate the uncertainty by a least-squares fit to a Lorentzian.  Let $\model_\beta(f)$ be the model function for the Lorentzian where $\beta = (\fc, \gamma)$ are the model parameters, $\fc$ the center frequency, and $\gamma$ the linewidth parameter. The variance of the estimated parameters $\hat{\beta}_i$ ($i=1,2$) can be estimated via the covariance matrix,
\begin{align}
\Sigma
  = \left(J^TJ\right)^{-1}
    \sigma_\mr{res}^2 \ ,
\end{align}
where
\begin{align}
J &= \left(\frac{\partial \model_{\hat{\beta}}\left(f_j\right)}{\partial\beta_i}\right)_{j,i}
\end{align}
is the Jacobian matrix of the model function at the estimated parameter values for the measured frequencies $f_j$ and
\begin{align}
\sigma_\mr{res}^2 &= \frac{1}{N-2} \sum_{j=1}^N{\left(\model_j-\model_{\hat{\beta}}(f_j)\right)^2}
\end{align}
is the variance of the residuals (noise variance), respectively.  $N=T/\ts$ is the number of samples and $\ts$ is the sampling time. The variances of the individual model parameters are then the diagonal elements of the covariance matrix. Thus, the uncertainty in the estimated center frequency is given by the square root of the first diagonal element of the covariance matrix,
\begin{align}
\sigma_{\fc} = \sqrt{\Sigma_{11}} =  \sqrt{\left[\left(J^TJ\right)^{-1}\right]_{11}\sigma_\mr{res}^2} \ .
\end{align}
Since the SNR saturates for peaks whose intrinsic linewidth is well resolved ($\Df>\df$), we know that the variance of the residuals stays constant for longer measurement times (see Eq.~(\ref{eq:snrEx})). However, the number of resources (number of frequency points in the spectrum $N$) used for the fit increases linearly together with the measurement time $T$. 

To find the scaling of the uncertainty, we have to evaluate the first diagonal element of $\left(J^TJ\right)^{-1}$ in the limit of a well resolved linewidth, 
\begin{align}
\left(J^TJ\right)^{-1} &=
\begin{pmatrix}
\sum_{j=1}^N\left(\frac{\partial}{\partial \fc} \model_{\hat{\beta}}(f_j)\right)^2 & \sum_{j=1}^N\left(\frac{\partial}{\partial \gamma} \model_{\hat{\beta}}(f_j)\right)\left(\frac{\partial}{\partial \fc} \model_{\hat{\beta}}(f_i)\right) \\
\sum_{j=1}^N\left(\frac{\partial}{\partial \gamma} \model_{\hat{\beta}}(f_j)\right)\left(\frac{\partial}{\partial \fc} \model_{\hat{\beta}}(f_j)\right) & \sum_{j=1}^N\left(\frac{\partial}{\partial \gamma} \model_{\hat{\beta}}(f_j)\right)^2  
\end{pmatrix}^{-1} \ .
\end{align}
Since we assume $\Df>\df$, we can approximate the sums with their corresponding integrals.  For the first diagonal entry, we find
\begin{align}
\left[\left(J^TJ\right)^{-1}\right]_{11} = \frac{f_s}{2N}\underbrace{\left(\int_0^{\frac{f_s}{2}}\left(\frac{\partial}{\partial \fc} \model_{\hat{\beta}}(f)\right)^2\mathrm{d}f\right)^{-1}}_{= u({\hat{\beta}})} =  \frac{f_s}{2N}\times u({\hat{\beta}}) \ ,
\label{eq:integralApprox}
\end{align}
where the factor $u({\hat{\beta}})$ is assumed to be approximately constant, since the estimated parameters $\hat{\beta}$ are at a minimum. Hence, the uncertainty of the center frequency scales as
\begin{align}
\sigma_{\fc} \propto N^{-0.5} \propto T^{-0.5}.
\end{align}
In the situation where the intrinsic linewidth is smaller than the frequency resolution, $\Df<\df$, the approximation in Eq.~(\ref{eq:integralApprox}) is not valid.  Furthermore, the number of resources for fitting the center frequency is not increasing with longer measurement times. We are given only three points in the spectrum to estimate the center frequency value: The point in the spectrum carrying most of the power and the two neighboring points left and right to that center peak. However, the SNR increases linearly and thus the relative noise variance decreases linearly. Therefore, the $T^{-1}$ scaling of the frequency resolution given by the Fourier transformation is boosted by $T^{-0.5}$ due to the SNR scaling, resulting in a overall uncertainty scaling of
\begin{align}
\sigma_{\fc}  \propto T^{-1.5} \ .
\end{align}

%%%%%
\newpage
\section*{Supplementary Text 3: Details of signal-to-noise ratio derivation}

This section provides the theoretical background for the signal-to-noise ratio (SNR) presented in Eqs.~(4-6) of the main manuscript. We repeat here Eq.~(4) for reference,
\begin{align}
\SNR \approx \frac{Y_j}{\sigma_Y}.\label{eq:SNR}
\end{align}
$Y_j$ is the height of a signal peak at frequency $f_j$ in the power spectrum, and $\sigma_Y$ is the standard deviation of the baseline noise evaluated in a frequency range where no signal is present.  In the following we calculate the expectation values for $Y_j$ and $\sigma_Y$ based on Eqs. (2,3) of the main manuscript.  This will lead us to Eq.~(6) of the main manuscript.

In a first step, we calculate the expected signal that appears in the power spectrum for a time trace $\{ y_k \} _{k=0}^{N-1}$ of photon counts, where $y_k$ was sampled at times $t_k=k\ts$, $\ts$ is the sampling period and $N$ is the number of samples.  To compute the power spectrum we first perform a discrete Fourier transform (DFT) of the time trace and then calculate the absolute square of the individual components.
The individual components of the power spectrum are given by
\begin{align}
Y_j = \left|\hat{y}_j \right|^2= \left|\sum_{k=1}^{N}{y_k e^{-2 \pi i k j }}\right|^2 \ .
\end{align}
Note that our definition of the DFT does not include any normalization by the number of points, \ie, we do not normalize the DFT by $N$ or $\sqrt{N}$.  Therefore, the power in each component grows with the square of $N$. The expected power in component $Y_j$ is
\begin{align}
\mathrm{E}\left[Y_j\right] &= \sum_{k=0,l=0}^{N-1}{\mathrm{E}\left[y_k y_l\right] e^{-2 \pi i k j }e^{2 \pi i l j }}\\
&= \sum_{k=0,l=0}^{N-1}{\mathrm{E}\left[y_k\right] \mathrm{E}\left[y_l\right] e^{-2 \pi i k j }e^{2 \pi i l j }}+
\sum_{k=0,l=0}^{N-1}{\mathrm{cov}\left[y_k,y_l\right] e^{-2 \pi i k j }e^{2 \pi i l j }}\\
&= \underbrace{\left|\mathrm{E}\left[\hat{y}_j \right]\right|^2}_{\text{signal contribution}}
 + \underbrace{\sum_{k=0}^{N-1}{\mathrm{var}(y_k)}}_{\text{noise contribution}}.
\end{align}
The last equation holds because any two samples at different times are independent.  We find that the power contained in component $Y_j$ is the sum of two contributions, one by the a.c. signal (first term) and one by the noise (second term).
The expected noise contribution is given by 
\begin{align}
\sum_{k=0}^{N-1}{\mathrm{var}(y_k)} = N \sigma_y^2  \label{eq:noiseA}
\end{align}
where $\sigma_y^2 = \mathrm{var}(y_k)$ for any $k$. This term represent the noise floor in the spectrum that is unrelated to the a.c. signal. In particular, as we will show below, this noise floor is present at any frequency, and has no frequency dependence, \ie, the noise floor is flat.

Next, we determine the noise entering the SNR.  The noise is given by the standard deviation of $Y_j$,
\begin{align}
\sigma_Y = \mathrm{std}\left(Y_j \right)\ .
\end{align}
To calculate $\sigma_Y$, we consider a power spectrum of a random stationary process without any additional a.c. signal, \ie, we assume $p(t_k) = 0.5$. We indicate quantities corresponding to this signal by a tilde, e.g. $\{\tilde{y}_k\}$ would be its time trace of measurement outcomes. Let $\tilde{R}_j$ and $\tilde{I}_j$ be the real and imaginary parts of $\hat{\tilde{y}}_j = \tilde{R}_j+i \tilde{I}_j$. Then, for large $N$, the central limit theorem implies that $\tilde{R}_j$ and $\tilde{I}_j$ have a normal distribution with zero mean and variance $\tilde{\sigma}^2$, where $\tilde{\sigma}^2$ is unknown. $\tilde{R}_j$ and $\tilde{I}_j$ are independent and identically distributed for $j < \frac{N}{2}$. We now write the power as sum of the power in the two quadratures
\begin{align}
\tilde{Y}_j=\left|\hat{\tilde{y}}_j \right|^2 = \tilde{R}_j^2 + \tilde{I}_j^2.
\end{align}
Furthermore, we realize that
\begin{align}
\frac{1}{\tilde{\sigma}^2}\tilde{Y}_j = \frac{1}{\tilde{\sigma}^2}\tilde{R}_j^2 + \frac{1}{\tilde{\sigma}^2}\tilde{I}_j^2 \sim \chi^2(2)
\end{align}
is $\chi$-square distributed with two degrees of freedom. This implies that 
\begin{align}
\mathrm{std}\left(\tilde{Y}_j \right) = \mathrm{E}\left[\tilde{Y}_j\right] \ ,
\end{align}
\ie, the standard deviation of the noise floor equals the expectation value of the noise floor. Using Parseval's theorem and again omitting a static offset, the standard deviation of the noise floor can also be related to the noise in the time trace,
\begin{align}
\mathrm{std}\left(\tilde{Y}_j \right) = \mathrm{E}\left[\tilde{Y}_j\right] = \sum_{k=0}^{N-1}{\mathrm{var}(\tilde{y}_k)} \approx \sum_{k=0}^{N-1}{\mathrm{var}(y_k)} = N \sigma_{y}^2 \label{eq:noiseB}\ .
%\mathrm{E}\left[\tilde{Y}_j\right] = \sum_{k=0}^{N-1}{\mathrm{var}(\tilde{y}_k)} \approx \sum_{k=0}^{N-1}{\mathrm{var}(y_k)} = N \sigma_{y}^2 \label{eq:noiseB}\ .
\end{align}
%
%As expected, the noise contribution to $\mathrm{E}\left[Y_j\right]$ (\ref{eq:noiseA}) and the noise floor of the measurement (\ref{eq:noiseB}) are the same. Furthermore, we can now evaluate the noise in our measurements either by computing the variance of the noise floor in the power spectrum or by computing the variance of the photon counts.
%
To obtain the SNR, we divide the expected signal $\left|\mathrm{E}\left[\hat{y}_j \right]\right|^2$ by the noise standard deviation $N \sigma_{y}^2$,
\begin{align}
\SNR
  &= \frac{\left|\mathrm{E}\left[\hat{y}_j \right]\right|^2}{N \sigma_{y}^2}
	 = \frac{\mathrm{E}\left[Y_j \right] - N \sigma_{y}^2}{N \sigma_{y}^2}
   = \frac{\mathrm{E}\left[Y_j \right]}{N\sigma_y^2}-1
	 \approx \frac{\mathrm{E}\left[Y_j \right]}{N\sigma_y^2}
	 = \frac{\mathrm{E}\left[Y_j \right]}{\sigma_Y}\ ,
\end{align}
where the approximation is for large SNR, which was the case in our measurements.  This corresponds to Eq. (4) in the main manuscript.
%We notice here, that $\mathrm{std}\left(\tilde{Y}_j \right)$ equals $\sigma_Y$ in a frequency range where no signal is present. 

%%%

In a next step, we explicitly calculate the SNR that applies to our detection scheme.  The expected a.c. signal contribution is given by
\begin{align}
\left|\mathrm{E}\left[\hat{y}_j \right]\right|^2 
&= \left|\mathrm{E}\left[\sum_{k=0}^{N-1}{y_k e^{-2 \pi i k j }} \right]\right|^2\\
&= \left|\sum_{k=0}^{N-1}{\mathrm{E}\left[y_k\right] e^{-2 \pi i k j }} \right|^2\\
&= \left|\sum_{k=0}^{N-1}{\Cr\epsilon \ p(t_k) e^{-2 \pi i k j }} \right|^2 \label{eq:omittedOffset}\\ 
&= \left|\sum_{k=0}^{N-1}{\Cr\epsilon \ \frac{1}{2}\phi_k e^{-2 \pi i k j }} \right|^2 \label{eq:omittedOffset2}\\
&= \frac{1}{4} \left(\Cr\epsilon\right)^2 \left|\hat{\phi}_j \right|^2, \label{eq:signalIdeal}
\end{align}
where we have omitted static offsets in Eq.~(\ref{eq:omittedOffset}) and Eq.~(\ref{eq:omittedOffset2}) that only contribute to the $j=0$ component. Here, $\hat{\phi}_j$ are Fourier components of the phases $\phi_k$ acquired by the quantum lock-in instances. We find that the signal power is proportional to the square of the readout gain $\Cr$ and the square of the optical contrast $\epsilon$. 
%Using Eq.~(\ref{eq:linearTransProb}).%, we find the signal power to be
%
%\begin{align}
%\left|\mathrm{E}\left[\hat{y}_j \right]\right|^2  &= \frac{1}{4} \left(\Cr\epsilon\right)^2 \left(\frac{\phi_{\mathrm{max}}}{2N} \right)^2\ .
%\end{align}

The noise variance $\sigma_y^2$ is calculated from Eq.~(3) in the main manuscript, which includes two random processes, a Bernoulli process associated with the quantum state projection and a Poisson process associated with the photon shot noise.  Both processes contribute to the noise variance.  To compute the contribution by the Bernoulli process, we assume that the state probability $p(t_k)$ is oscillating closely around the $p=0.5$ bias point.  Then, the variance in $p$ is $\frac14$ and the corresponding variance in $y_k$ is $\frac14(\Cr\epsilon)^2$.  The contribution by the Poisson process has a variance that is equal to the mean of $y_k$, which is $\frac12C(1-\epsilon/2)$.  The total noise variance $\sigma_y^2$ is then
\begin{align}
\sigma_y^2 = \frac14\Cr^2\epsilon^2 + C\left(1-\frac{\epsilon}{2}\right) \ .
\end{align}
This yields the explicit expression for the SNR,
\begin{align}
\SNR
  &= \frac{\frac{1}{4}(\Cr\epsilon)^2 |\hat{\phi}|^2}{N\left[\frac{1}{4}(\Cr\epsilon)^2+\Cr(1-\frac{\epsilon}{2})\right]}
   = \frac{\frac{1}{4}(\Cr\epsilon)^2\ N }{\frac{1}{4}(\Cr\epsilon)^2+\Cr(1-\frac{\epsilon}{2})} |\hat{\phi}_j/N|^2 \ . \label{eq:snrEx}
\end{align}
This SNR applies to a general Fourier component of the spectrum $Y_j$.  If the spectrum has very narrow peaks, such that the entire signal power is concentrated in a single Fourier component $Y_j$, we have
\begin{align}
|\hat{\phi}_j/N|^2 =\frac{1}{4}\phimax^2 \ .
\end{align}
This situation corresponds to the case where the intrinsic linewidth $\Df$ of the signal is smaller than the frequency resolution $\df$, which is the typical situation for our experiments.  In this situation, the SNR is
\begin{align}
\SNR = \frac{\frac{1}{16}(\Cr\epsilon)^2\ N \phimax^2}{\frac{1}{4}(\Cr\epsilon)^2+\Cr(1-\frac{\epsilon}{2})} \ . \label{eq:snrIdeal}
\end{align}
This is Eq. (6) of the main manuscript.  If the signal has only one frequency component and the lock-in is tuned to that frequency, $\phimax = 2\ta\Omega/\pi$ (see Eq. (\ref{eq:phimax})).

%where we now normalized the power of the phase  to get the average power per sample ($|\hat{\phi}_j/N|^2$). This is useful, because this quantity is independent of the length of the measured time trace $\{y_k \}$. In particular, if the entire signal power is concentrated in one Fourier component $\hat{\phi}_j$, we have $|\hat{\phi}_j/N|^2 =  \frac{1}{2}\phi^2_{\mathrm{rms}}$ (and $|\hat{\phi}_j/N|^2 =\frac{1}{4}\phi_{\mr{max}}$). Furthermore, in this case, the SNR value is proportional to the square of the ac signal amplitude $\Omega$, since $\phi_{\mathrm{rms}} \propto \Omega$ (and $\phi_{\mr{max}} \propto \Omega$).

Eq. (\ref{eq:snrIdeal}) represents the SNR for an ideal read-out process.  In our experiments, the read out was compromised by the limited robustness of the nuclear \NN quantum memory.  With each quantum non-demolition (QND) measurement of the memory qubit, there is a finite chance of depolarizing the qubit and losing the stored information.  Although this effect is rather weak in our case, with a spin flip probability per QND measurement of $\Gamma \lesssim 0.1\%$, it needs to be considered for large QND repetitions $n$.

Because the depolarization probability is small, we can restrict ourselves to the following two cases:  Either there are zero nuclear spin flips during readout, or there is a non-zero number of spin flips.  In the first case, the original state is detected during the entire readout and we gain the correct information about the transition probability $p(t_k)$.  By contrast, in the second case, the information is lost along the readout process.  The probability distribution of photon counts $y_k$ is given by
\begin{align}
f\left(y_k|p\left(t_k\right)\right) = e^{-\Gamma n}\ f\left(y_k|\ p\left(t_k\right) \wedge \mathrm{no\ flip}\ \right)+ (1-e^{-\Gamma n})\tilde{f}\left(y_k| \ \mathrm{flip} \ \right).
\end{align}
where $n$ is the number of QND repetitions.
We note that the probability distribution function $\tilde{f}\left(y_k| \ \mathrm{flip} \ \right)$ in the case of a non-zero number of spin flips is independent of the transition probability $p(t_k)$. Similar to Eq. (\ref{eq:signalIdeal}), we can compute the expected power in the spectrum as
\begin{align}
\left|\mathrm{E}\left[\hat{y}_j \right]\right|^2
&=\frac14(\Cr\epsilon)^2\ e^{-2\Gamma n}\  \left|\hat{\phi}_j\right|^2 ,
\end{align}
This yields a modified SNR given by
\begin{align}
%\SNR = \frac{\frac14(\Cr\epsilon)^2\ N e^{-2\Gamma n}}{\frac{1}{4}(\Cr\epsilon)^2+\Cr(1-\frac{\epsilon}{2})} \left|\hat{\phi}_j/N \right|^2
\SNR = \frac{\frac{1}{16}(\Cr\epsilon)^2\ N e^{-2\Gamma n}\phimax^2}{\frac{1}{4}(\Cr\epsilon)^2+\Cr(1-\frac{\epsilon}{2})} \ .
\end{align}
We have used this equation to fit the data in Fig.~4 of the main text and to extract values for the readout gain $\Cr$, the optical contrast $\epsilon$ and the spin flip rate $\Gamma$.

%%%%%%%%%%%%%%%%%%%%%

\newpage

\section*{Supplementary Text 4: Details of the compressive sampling protocol}
\label{sec:unfolding}

Due to the nature of the lock-in measurement and the long sampling period $\ts$, we acquire sample records $\ytrace$ at rates $\fs=1/\ts$ that are far below the Nyquist rate for the a.c. signals.  This means that our continuous sampling strategy enables an arbitrarily fine frequency resolution only in a narrow bandwidth, and does not reproduce the absolute signal frequency.  To reconcile the absolute signal frequency, we record the same a.c. signal several times with slightly different sampling rates $\fs$.  In our detection scheme, we can adjust $\fs$ by adding a small extra delay to the delay time $\td$.  We then reconstruct the wideband spectrum based on a compressive sampling (CS) technique \cite{donoho06}.

CS refers to the idea that certain types of signals, more exactly signals which are sparse in some basis, can be reconstructed out of a small number of partial measurements. Specifically, suppose that we have a discrete number of samples $M$ of a signal $x \left( t \right)$ with $0<t<T$.  Then $x \left( t \right)$ can be represented by a set of basis functions $\varphi_{k}\left(t\right)$, for example the Fourier basis $\varphi_{k}\left(t\right) = e^{i2\pi f_k t}$, as
\begin{align}
x(t) = \sum_{k=0}^{M-1} X_{k} \varphi_{k}\left(t\right) \ .
\label{eq:fourier_series}
\end{align}
If only a few coefficients $X_{k}$ are significantly non-zero, then the signal $x \left( t \right)$ is considered \textit{sparse} and its reconstruction from a set of measurements acquired at sub-Nyquist rates becomes an optimization problem \cite{candes06,donoho06}.

Let us assume that the signal is represented by the vector $\vec{x}=\left\{x_{0},\ldots,x_{m},\ldots,x_{M}\right\}$ and that the samples of $\vec{x}$ are acquired at or above a relevant Nyquist rate.  Then the discrete wideband spectrum $\vec{X}=\left\{X_{0},\ldots,X_{k},\ldots,X_{M}\right\}$ of $\vec{x}$ is given by the set of coefficients $X_{k}$,
\begin{align}
X_{k} = \sum_{m=0}^{M-1} x_{m}\exp\left(-i2\pi k\frac{m}{M}\right) \ ,
\label{eq:dft}
\end{align}
The process of measuring an undersampled spectrum $\vec{Y}_{i}$ of $\vec{X}$ can be viewed as the action of a sampling matrix \textbf{$\Phi_{i}$} on the target spectrum $\vec{X}$. If each partial measurement $\vec{Y}_{i}$ consists of $N_{i}$ samples, then each \textbf{$\Phi_{i}$} has dimensions $N_{i} \times M$. It has been shown that $\vec{X}$ can be recovered by using $p\approx s\mathcal{O}\left(\log (M)\right)$ partial measurements where $s \ll M$ indicates the sparsity of $\vec{X}$, \ie the number of significantly non-zero coefficients of $\vec{X}$ \cite{polo09}.
The problem can be written as
\begin{align}
\bm{\vec{Y}} = 
\left(\begin{array}{c}
    \vec{Y}_{1} \\
    \vdots \\
    \vec{Y}_{p}
\end{array}\right)
 = 
\left(\begin{array}{c}
    \Phi_{1} \\
    \vdots \\
    \Phi_{p}
\end{array}\right)\vec{X}
= \bm{\Phi} \vec{X},
\label{eq:compressiveproblem}
\end{align}
where $\bm{\vec{Y}}$ is a vector that contains the $p$ undersampled spectra $\vec{Y}_{i}$ and $\bm{\Phi}$ a vector that contains the $p$ sampling matrices \textbf{$\Phi_{i}$} of dimensions $N_{i}\times M$ each. Thus, reconstruction of $\vec{X}$ out of the set of $\vec{Y}_{i}$ requires appropriate construction of the sampling matrices \textbf{$\Phi_{i}$} and dedicated algorithms. The problem has motivated research in the context of wideband spectrum sensing \cite{nader11}, where the idea is to achieve awareness of spectral opportunities, \ie, to detect and fill licensed but unused portions of the electromagnetic spectrum at minimal computational cost. For this purpose, approaches like $l_{1}$ minimization or greedy pursuit algorithms have been studied \cite{needell10,sun12}.

In our experiment, we implemented a compressive sensing scheme where the $\vec{Y}_i$ represent undersampled spectra acquired at slightly different sampling rates $\fsi$.  We varied the sampling rates by adding small extra delays to the delay time $\td$.  We then constructed a sampling matrix $\Phi_i$ for each spectrum $\vec{Y}_i$ following Refs. \cite{sun12,needell10} and solved the linear system (Eq.~\ref{eq:compressiveproblem}) by a non-negative linear least-squares solver ({\texttt{lsqnonneg}}) in Matlab.

To successfully reconstruct $\vec{X}$, suitable sampling matrices \textbf{$\Phi_{i}$} need to be chosen.  The reason is that we want to avoid that two matrices \textbf{$\Phi_{i}$} and \textbf{$\Phi_{j}$} map the same components of $\vec{X}$ into different undersampled spectra $\vec{Y}_{i}$ and $\vec{Y}_{j}$.  To avoid such a situation, the matrices should be chosen as orthogonal or maximally incoherent as possible.
The coherence $\mu$ of the sampling matrices is obtained via the inner product of their columns
\begin{align}
\mu = \max_{i\neq j\in\left[1,M\right]}\left|\left\langle \phi_{i},\phi_{j}\right\rangle\right|
\label{eq:matrix_coherence}
\end{align}
where $\phi_{i}$ denotes a $l_{2}$-normalized column of the matrix $\bm{\Phi}$. $\mu$ is a measure of the orthogonality of the sampling matrices, and under appropriate construction equals $1/p$ \cite{sun12}. In such case, the spectrum $\vec{X}$ can be exactly reconstructed if $p>2s-1$ \cite{donoho06}.

We minimized the coherence of our sampling matrices by choosing random delay times $\td$, which in turn determine $\fsi$ and therefore $N_i=T\fsi$. Furthermore, since the effective total measurement durations $T_i\approx T$ were not identical (due to rounding requirements of the pulse generator), the frequency resolution for a  measured $\vec{Y}_i$ and the sought after spectrum $\vec{X}$ are not exactly the same.  We therefore interpolated the elements of the sampling matrices $\Phi_i$, which would ideally be $\in\{0,1\}$, to fractional values $\in [0...1]$, as
\begin{align*}
\Phi_{i}\left[n',m'\right] &= \frac{N_i}{M} \sum_{l=-\infty}^{\infty} w_{nm} \left(\delta\left(m-\left\lfloor (n+l N_i)\frac{T}{T_i} \right\rfloor \right) + \delta\left(m-\left\lceil (n+l N_i)\frac{T}{T_i} \right\rceil \right)\right)\\
n' &= n + \left\lfloor\frac{N_i}{2}\right\rfloor + 1\\
m' &= m + \left\lfloor\frac{M}{2}\right\rfloor + 1\\
w_{nm} &= \left|1 - \left(m - (n+l N_i)\frac{T}{T_i}\right)\right|
\end{align*}
where $\left\lfloor a \right\rfloor$ is the floor function and $\left\lceil a \right\rceil$ is the ceil function. $\delta\left(a\right)$ denotes the Kronecker delta function and $\left|a\right|$ the absolute value. To minimize computational costs, we only reconstructed the portions of the spectrum falling within the CPMG filter windows, \ie, we constructed sparse matrices $\Phi_{i}$ with zeros everywhere outside the spectral portions of interest.

%%%%%%%%%%%%%%%%%%%%%%%%%%%%%%%%%%%%%%%%%%%%%%%%%%%%%%%%%%%%%%%%%%%%%%%%%%%%%%%%%%%%%%%%%%%%%%%%%%%%%%%%%%%%%

%%% sensing sequence
\newpage{}
\section*{Supplementary Figure 1}

%%% CPMG sequence
\newpage{}
\section*{Supplementary Figure 2}

%%% Spectra for determining linewidth and uncertainty as a function of T
\newpage{}
\section*{Supplementary Figure 3}

\begin{figure}[h!]
\centering
\includegraphics[width=0.80\textwidth]{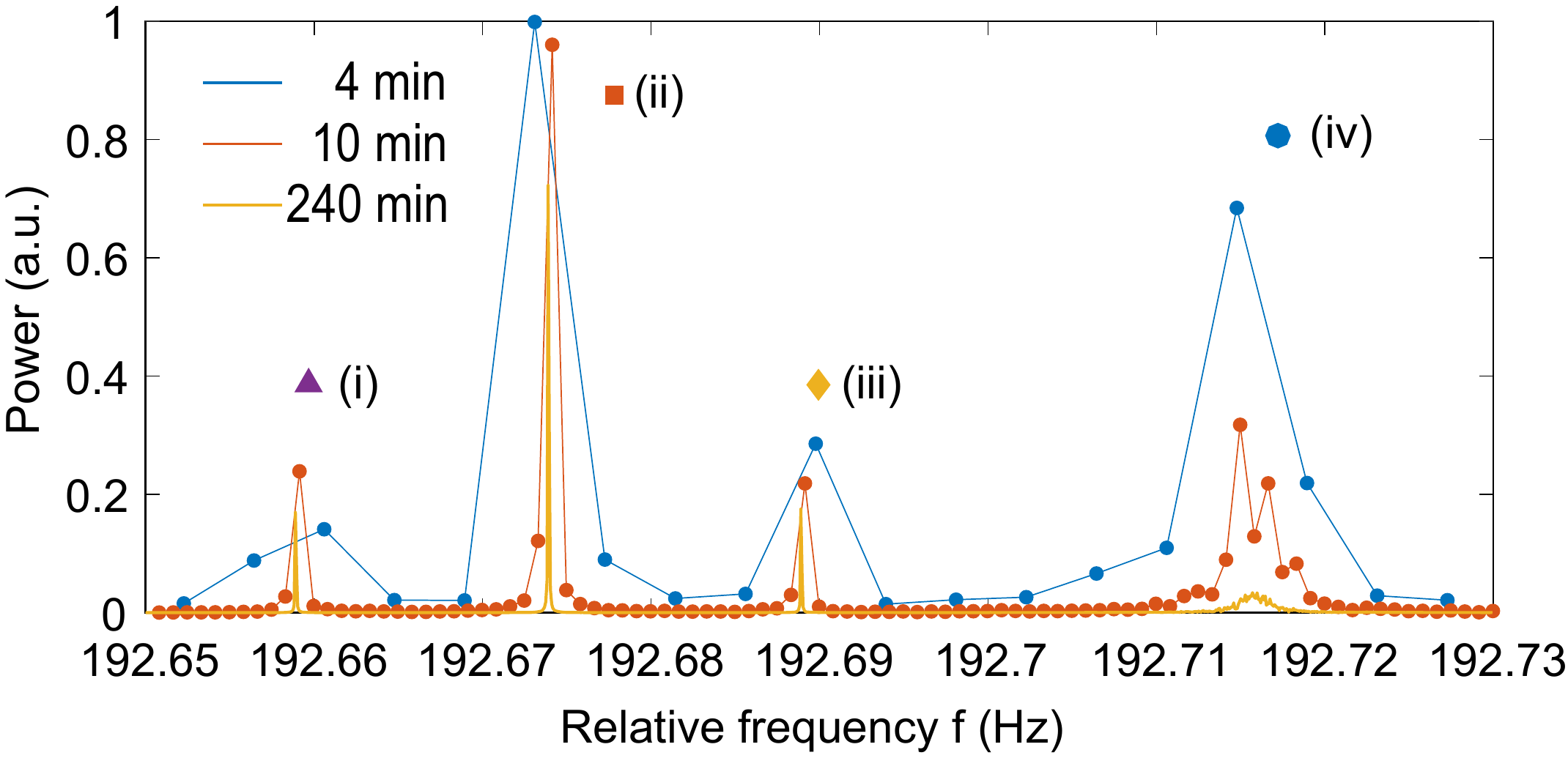}
\caption{
Power spectrum of a.c. signals for $T=4\unit{min}$, $10\unit{min}$ and $240\unit{min}$.  Peaks (i-iii) originate from coherent signals at $\fc=1.2\unit{MHz}$ and $\fc\pm 15\unit{mHz}$ produced by amplitude modulation.
Signal (iv) with frequency $\fc+40\unit{mHz}$ originates from a frequency modulated (FM) signal with an artificial line broadening of $\Df = 0.76\unit{mHz}$.
}
	\label{fig:Linewidth}
\end{figure}

%%% Bessel plots
\newpage{}
\section*{Supplementary Figure 4}

\begin{figure}[h!]
\centering
\includegraphics[width=0.99\textwidth]{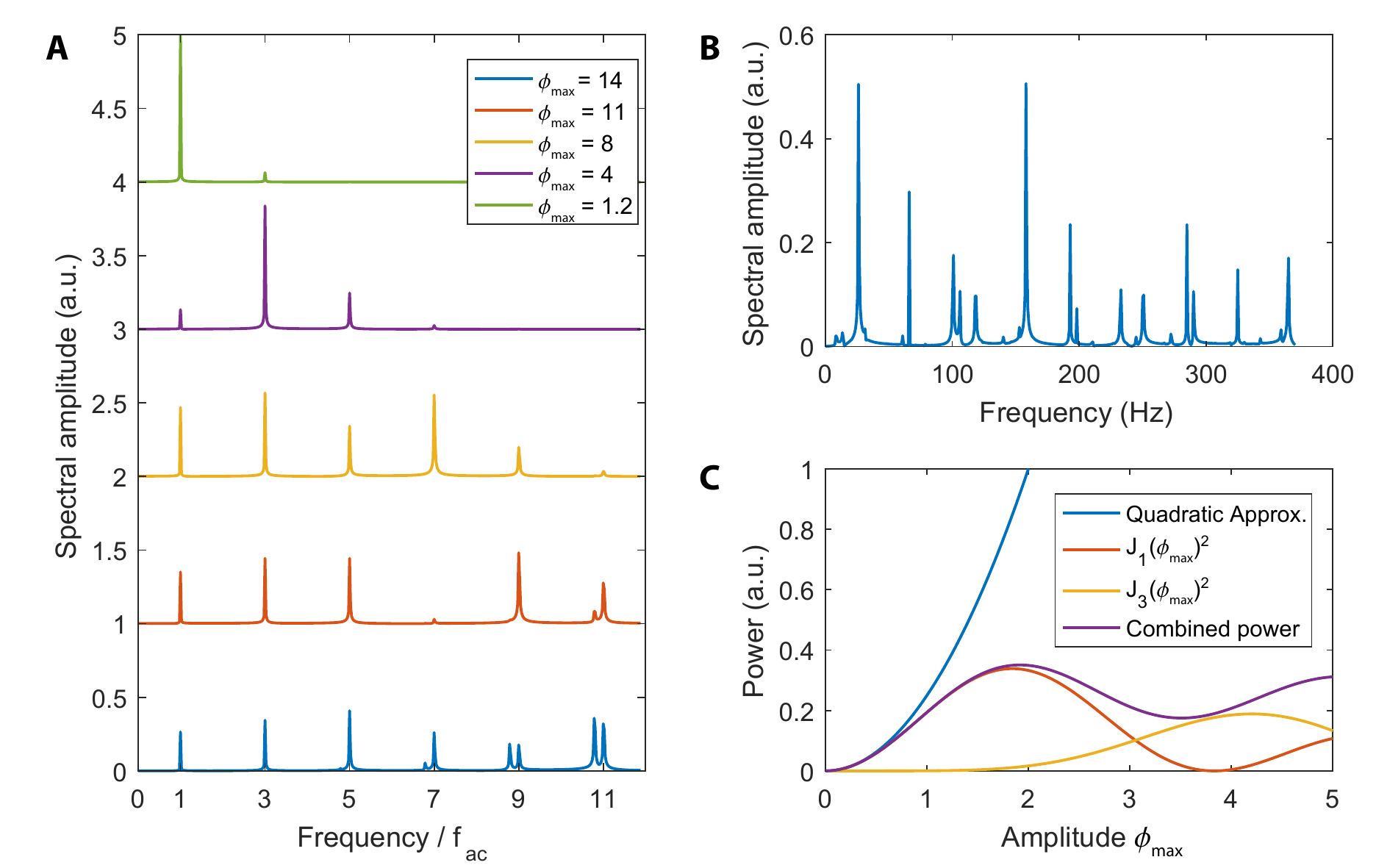}
\caption{
(\textbf{A}) Simulated spectra based on formula given in Eq.~(\ref{eq:nonLinearSig}) for different signal amplitudes $\phimax$.  Odd harmonics of the signal frequency $\fac$ are observed as $\phimax$ exceeds the linear regime of the sensor ($\phimax\gtrsim 1$).  The number of harmonics increases with the signal amplitude $\phimax$, eventually leading to spectral folding.
(\textbf{B}) Simulated spectrum of a signal with two frequency components: $\fac = 400.75$~kHz and  $\fac' = 401.75$~kHz sampled at a rate of $\fs = 742.1$~Hz.  The signal amplitudes are $\phi_\mr{max,1}=3$ and $\phi_\mr{max,2}=2$, respectively.  The spectrum shows harmonics as well as frequency mixing of the two fundamental frequencies.  Spectral folding further complicates the interpretation of the spectrum.
(\textbf{C}) Peak height in the power spectrum as a function of $\phimax$ for a signal with frequency $\fac$.  The blue curve is given by $\phirms^2/4$ and represents the linear regime where $J_1(\phimax)\approx\phimax/2$.  We have used this approximation in our experiments.  The red and yellow curves show the first and third Bessel function corresponding to the amplitudes of the $\fac$ and $3\fac$ harmonics, respectively.  The purple curve shows the total power of all harmonics, $\sum^{\infty}_{k=0} J_{2k+1}^2(\phimax)$, which approaches $0.25$ as $\phimax\rightarrow\infty$.
}
	\label{fig:besselPlots}
\end{figure}

%%% Harmonics
\newpage{}
\section*{Supplementary Figure 5}

\begin{figure}[h!]
\centering
\includegraphics[width=0.99\textwidth]{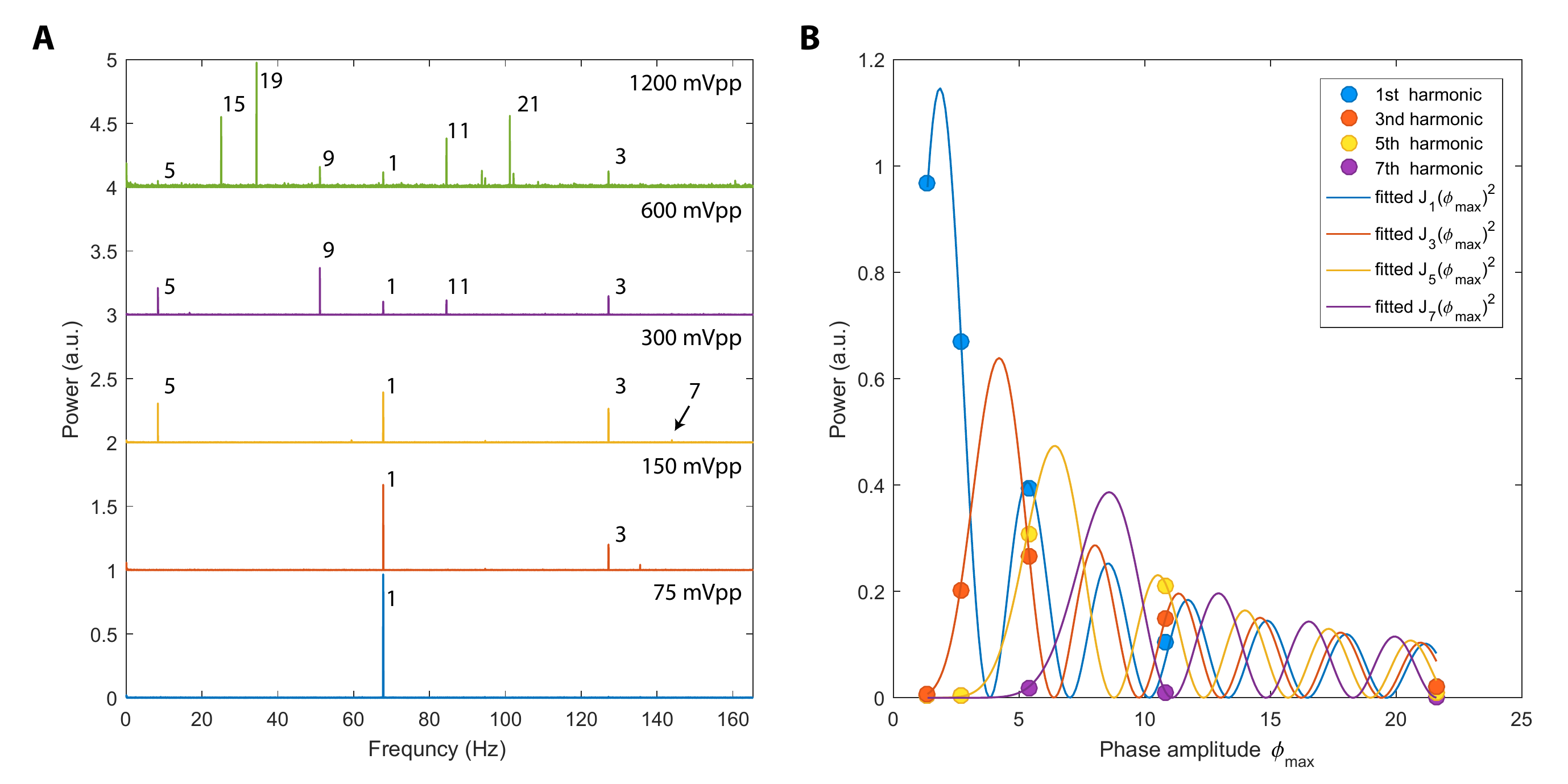}
\caption{
(\textbf{A}) Experimental spectra of a single tone a.c. signal with $\fac=1.202254655\unit{MHz}$ for five different amplitude settings on the external signal generator.  The amplitude settings are stated with each plot.  The spectra are vertically shifted by one unit for clarity.  All spectra use the same vertical scale, except for the top spectrum which is magnified $6\times$.  Labels identify the different signal harmonics as discussed with Eq. (\ref{eq:nonLinearSig}).  
(\textbf{B})  Fitted peak heights for the first four harmonics at $\fac$, $3\fac$, $5\fac$ and $7\fac$ as a function of the phase amplitude $\phimax$.  The data are in excellent agreement with Eq. (\ref{eq:nonLinearSig}).
}
	\label{fig:harmonics}
\end{figure}

%%%%%%%%%%%%%%%%%%%%%%%%%%%%%%%%%%%%%%%%%%%%%%%%%%%%%%%%%%%%%%%%%%%%%%%%%%%%%%%%%%%%%%%%%%%%%%%%%%%%%%%%%%%%

\newpage
\section*{Supplementary References}

%\setcounter{enumiv}{9}
%\bibliography{library}
%\bibliography{C:/Christian/ETH/labview/library/library}

%\begin{comment}
%\begin{thebibliography}{15}
%\makeatletter
%\addtocounter{\@listctr}{8}
%\makeatother
%\end{thebibliography}
%\end{comment}